\documentstyle[epsf]{article}

\newfont{\ssr}{cmss10}
\newfont{\smssr}{cmss10 scaled 800}
\newcounter{subequation}[equation]

\makeatletter

\def\@aabuffer{}
\def\author #1{\expandafter\def\expandafter\@aabuffer\expandafter
{\@aabuffer \small\rm      #1\relax \par}}
\def\address#1{\expandafter\def\expandafter\@aabuffer\expandafter
{\@aabuffer \small\it #1\relax \par\vspace{1em}}}

\def\maketitle{
\begin{center}
   {\bf \@title \par}
   \vskip 2em                      % Vertical space after title.
   \@aabuffer\relax
\end{center} \par
\gdef\@aabuffer{}
}

\def\abstracts#1{
\begin{center}
{\begin{minipage}{4.2truein}
                 \footnotesize
                 \parindent=0pt #1\par
                 \end{minipage}}\end{center}
                 \vskip 2em \par}

\fussy
\def\section{\@startsection {section}{1}{\z@}{-3.5ex plus -1ex minus
    -.2ex}{2.3ex plus .2ex}{\bf }}
\def\subsection{\@startsection{subsection}{2}{\z@}{-3.25ex plus -1ex minus
   -.2ex}{1.5ex plus .2ex}{\it }}

\def\thefootnote{\alph{footnote}}
\def\@makefnmark{{$\!^{\@thefnmark}$}}

\renewenvironment{thebibliography}[1]
	{\begin{list}{\arabic{enumi}.}
	{\usecounter{enumi}\setlength{\parsep}{0pt}
	 \setlength{\itemsep}{0pt}
         \settowidth
	{\labelwidth}{#1.}\sloppy}}{\end{list}}

\newcounter{arabiclistc}

\def\citen#1{%
\edef\@tempa{\@ignspaftercomma,#1, \@end, }% ignore spaces in parameter list
\edef\@tempa{\expandafter\@ignendcommas\@tempa\@end}%
\if@filesw \immediate \write \@auxout {\string \citation {\@tempa}}\fi
\@tempcntb\m@ne \let\@h@ld\relax \def\@citea{}%
\@for \@citeb:=\@tempa\do {\@cmpresscites}%
\@h@ld}
\def\@ignspaftercomma#1, {\ifx\@end#1\@empty\else
   #1,\expandafter\@ignspaftercomma\fi}
\def\@ignendcommas,#1,\@end{#1}
\def\@cmpresscites{%
 \expandafter\let \expandafter\@B@citeB \csname b@\@citeb \endcsname
 \ifx\@B@citeB\relax % undefined
    \@h@ld\@citea\@tempcntb\m@ne{\bf ?}%
    \@warning {Citation `\@citeb ' on page \thepage \space undefined}%
 \else%  defined
    \@tempcnta\@tempcntb \advance\@tempcnta\@ne
    \setbox\z@\hbox\bgroup % check if citation is a number:
    \ifnum0<0\@B@citeB \relax
       \egroup \@tempcntb\@B@citeB \relax
       \else \egroup \@tempcntb\m@ne \fi
    \ifnum\@tempcnta=\@tempcntb % Number follows previous--hold on to it
       \ifx\@h@ld\relax % first pair of successives
          \edef \@h@ld{\@citea\@B@citeB }%
       \else % compressible list of successives
          \edef\@h@ld{\hbox{--}\penalty\@highpenalty
            \@B@citeB }%
       \fi
    \else   %  non-successor--dump what's held and do this one
       \@h@ld\@citea\@B@citeB
       \let\@h@ld\relax
 \fi\fi%
 \def\@citea{,\penalty\@highpenalty\hskip.13em plus.1em minus.1em}%
}
\def\@citex[#1]#2{\@cite{\citen{#2}}{#1}}%
\def\@cite#1#2{\leavevmode\unskip
 \ifnum\lastpenalty=\z@\penalty\@highpenalty\fi%highpenalty before
     {\multiply\@highpenalty 3 $\!^{#1}$%      %triple-high penalties
     \if@tempswa,\penalty\@highpenalty\ #2\fi  %within and before note.
   }\spacefactor\@m}

\def\baselinestretch{1.0}
\ifx\selectfont\undefined
\@normalsize\else\let\glb@currsize=\relax\selectfont
\fi

\ifx\selectfont\undefined
\def\@singlespacing{%
\def\baselinestretch{1}\ifx\@currsize\normalsize\@normalsize\else\@currsize\fi%
}
\else
\def\@singlespacing{\def\baselinestretch{1}\let\glb@currsize=\relax\selectfont}
\fi

\long\def\@makecaption#1#2{
   \vskip 10pt
   \setbox\@tempboxa\hbox{\footnotesize #1: #2}
   \ifdim \wd\@tempboxa >\hsize   % IF longer than one line:
       \leftskip 0pt plus 1fil
       \rightskip 0pt plus -1fil
       \parfillskip 0pt plus 2fil
       \footnotesize #1: #2\par   %   THEN set as ordinary paragraph.
     \else                        %   ELSE  center.
       \hbox to\hsize{\hfil\box\@tempboxa\hfil}
   \fi}

\def\thesubequation{\theequation\@alph\c@subequation}
\def\@subeqnnum{{\rm (\thesubequation)}}
\def\slabel#1{\@bsphack\if@filesw {\let\thepage\relax
   \xdef\@gtempa{\write\@auxout{\string
      \newlabel{#1}{{\thesubequation}{\thepage}}}}}\@gtempa
   \if@nobreak \ifvmode\nobreak\fi\fi\fi\@esphack}
\def\subeqnarray{\stepcounter{equation}
\let\@currentlabel=\theequation\global\c@subequation\@ne
\global\@eqnswtrue
\global\@eqcnt\z@\tabskip\@centering\let\\=\@subeqncr
$$\halign to \displaywidth\bgroup\@eqnsel\hskip\@centering
  $\displaystyle\tabskip\z@{##}$&\global\@eqcnt\@ne
  \hskip 2\arraycolsep \hfil${##}$\hfil
  &\global\@eqcnt\tw@ \hskip 2\arraycolsep
  $\displaystyle\tabskip\z@{##}$\hfil
   \tabskip\@centering&\llap{##}\tabskip\z@\cr}
\def\endsubeqnarray{\@@subeqncr\egroup
                     $$\global\@ignoretrue}
\def\@subeqncr{{\ifnum0=`}\fi\@ifstar{\global\@eqpen\@M
    \@ysubeqncr}{\global\@eqpen\interdisplaylinepenalty \@ysubeqncr}}
\def\@ysubeqncr{\@ifnextchar [{\@xsubeqncr}{\@xsubeqncr[\z@]}}
\def\@xsubeqncr[#1]{\ifnum0=`{\fi}\@@subeqncr
   \noalign{\penalty\@eqpen\vskip\jot\vskip #1\relax}}
\def\@@subeqncr{\let\@tempa\relax
    \ifcase\@eqcnt \def\@tempa{& & &}\or \def\@tempa{& &}
      \else \def\@tempa{&}\fi
     \@tempa \if@eqnsw\@subeqnnum\refstepcounter{subequation}\fi
     \global\@eqnswtrue\global\@eqcnt\z@\cr}
\let\@ssubeqncr=\@subeqncr
\@namedef{subeqnarray*}{\def\@subeqncr{\nonumber\@ssubeqncr}\subeqnarray}
\@namedef{endsubeqnarray*}{\global\advance\c@equation\m@ne%
                           \nonumber\endsubeqnarray}

\@addtoreset{equation}{section}
\renewcommand{\theequation}{\thesection.\arabic{equation}}

\makeatother

\def\Journal#1#2#3#4{{#1} {\bf #2}, #3 (#4)}

\def\IJMPA{{\em Int. J. Mod. Phys.} A}
\def\NPB{{\em Nucl. Phys.} B}
\def\PLB{{\em Phys. Lett.}  B}
\def\PRL{\em Phys. Rev. Lett.}
\def\PRD{{\em Phys. Rev.} D}
\def\CQG{\em Class. Quantum Grav.}
\def\st{\scriptstyle}
\def\sst{\scriptscriptstyle}

\def\be{\begin{equation}}
\def\ee{\end{equation}}
\def\bea{\begin{eqnarray}}
\def\eea{\end{eqnarray}}
\def\ft#1#2{{\textstyle{{\scriptstyle #1}\over {\scriptstyle #2}}}}
\def\fft#1#2{{#1 \over #2}}
\def\ffrac#1#2{\leavevmode\kern.1em
\raise.5ex\hbox{\the\scriptfont0 #1}\kern-.1em/\kern-.15em
\lower.25ex\hbox{\the\scriptfont0 #2}}
\def\sffrac#1#2{\leavevmode\kern.1em
\raise.5ex\hbox{\the\scriptscriptfont0 #1}\kern-.1em/\kern-.15em
\lower.25ex\hbox{\the\scriptscriptfont0 #2}}
\def\gtlt{\mathrel{\raise4.5pt\hbox{\oalign{$\scriptstyle>$\crcr
$\scriptstyle<$}}}}
\def\cramp{\medmuskip = 2mu plus 1mu minus 2mu}

\def\crampest{\medmuskip = 1mu plus 1mu minus 1mu}
\def\uncramp{\medmuskip = 4mu plus 2mu minus 4mu}
\def\R{\rlap{\rm I}\mkern3mu{\rm R}}
\def\Z{\rlap{\ssr Z}\mkern3mu\hbox{\ssr Z}}

\def\im{{\rm i}}
\def\dalemb#1#2{{\vbox{\hrule height .#2pt
        \hbox{\vrule width.#2pt height#1pt \kern#1 pt
                \vrule width.#2 pt}
        \hrule height.#2 pt}}}
\def\square{\mathord{\dalemb{6.2}{6}\hbox{\hskip1pt}}}
\def\for{\lower4pt\hbox{$\st|$}}
\def\oneone{\rlap 1\mkern4mu{\rm l}}

\begin{document}
\renewcommand{\thefootnote}{\fnsymbol{footnote}}
\null
\vspace{-3cm}
\rightline{Imperial/TP/96--97/15}
\rightline{\tt hep-th/9701088}
\vspace{1cm}

\title{LECTURES ON SUPERGRAVITY $p$-BRANES\,\footnote[1]{Lectures given
at the ICTP Summer School in High Energy Physics and Cosmology,
Trieste, June 10--26, 1996.}}

\author{K.S. STELLE}

\address{The Blackett Laboratory,\\ Imperial College,\\
Prince Consort Road,\\ London SW7 2BZ, UK}

\maketitle\abstracts{
We review the properties of classical $p$-brane
solutions to supergravity theories, {\it i.e.}\ solutions that may be
interpreted as Poincar\'e-invariant hyperplanes in spacetime. Topics
covered include the distinction between elementary/electric and
solitonic/magnetic solutions, examples of singularity and global
structure, relations between mass densities, charge densities and the
preservation of unbroken supersymmetry, diagonal and vertical Kaluza-Klein
reduction families, Scherk-Schwarz reduction and domain walls, and the
classification of multiplicities using duality symmetries.}

\renewcommand{\thefootnote}{\alph{footnote}}
\section{Introduction}\label{sec:intro}

     Supergravity theories originally arose from the desire to include
supersymmetry into the framework of gravitational models, and this was
in the hope that the resulting models might solve some of the
outstanding difficulties of quantum gravity. One of these difficulties
was the ultraviolet problem, on which early enthusiasm for
supergravity's promise gave way to disenchantment when it became clear
that local supersymmetry is not in fact sufficient to tame the notorious
ultraviolet divergences.\footnote{For a review of  ultraviolet behavior
in supergravity theories, see Ref.\cite{hs}} Nonetheless, these theories
won much admiration for their beautiful mathematical structure, which is
due to the stringent constraints of their symmetries. These severely
restrict the possible terms that can occur in the Lagrangian. For the
maximal supergravity theories, there is simultaneously a great wealth of
fields present and at the same time an impossibility
of coupling any independent external field-theoretic ``matter.'' It was
only occasionally noticed in this early period that this impossibility of
coupling to matter fields does not, however, rule out coupling to
``relativistic objects'' such as black holes, strings and membranes. 

     Indeed, a striking fact that has now been clearly recognized about
supergravity theories is the degree to which they tell us precisely what
kinds of external ``matter'' they will tolerate. The possibilities
of such couplings may be learned in a fashion similar to the traditional
derivation of the Schwarzshild solution in General Relativity, first
searching for an isotropic solution in empty space, then considering
later how this may be matched onto an interior matter source. In the
case of supergravity theories, imposing the requirement that some part of
the original theory's supersymmetry be left unbroken leads to the class
of $p$-brane solutions that we shall review in this article.\footnote{The
present review is based largely upon
Refs.\cite{stainless,dilatonic,weyl,vertical,domain} and focuses on
classical solutions to supergravity theories without special regard to the
structure of source terms that would reside in $p$-brane worldvolume
actions. This restricted focus is made here for simplicity, as the
structure of such $p$-brane worldvolume actions is still incompletely
known. Earlier reviews of $p$-brane solutions including discussion of the
worldvolume action sources may be found in Refs.\cite{pkt,dkl,mjd} The
occurrence of such solutions as $D$-brane backgrounds in string theory has
been recently reviewed in Ref.\cite{jp}} It is one of the marvels of the
subject that this purely {\em classical} information from supergravity
theories is now thought to be capable of yielding nonperturbative
information on the superstring theories that we now see as the underlying
{\em quantum} formulations of supergravity.

     First, let us recall the way in which long-wavelength limits
of string theories yield effective spacetime gravity theories.
Consider, to begin with, the $\sigma$-model action~~\cite{ft} that
describes a bosonic string moving in a background ``condensate''
of its own massless modes ($g_{\sst MN}$, $A_{\sst MN}$, $\phi$):
\bea
\lefteqn{I={1\over 4\pi\alpha'}\int
d^2z\sqrt\gamma\,[\gamma^{ij}\partial_i x^{\sst M}\partial_j x^{\sst N}
g_{\sst MN}(x)}\hspace{2cm}\nonumber\\
&&+ \im\epsilon^{ij}\partial_ix^{\sst M}\partial_jx^{\sst
N}A_{\sst MN}(x) +
\alpha'R(\gamma)\phi(x)]\ .\label{isigma}
\eea
Every string theory contains a sector described by fields
($g_{\sst MN}$, $A_{\sst MN}$, $\phi$); these are the only fields that
couple directly to the string worldsheet. In superstring theories,
this sector is called the Neveu-Schwarz/Neveu-Schwarz (NS-NS) sector.

     The $\sigma$-model action (\ref{isigma}) is classically
invariant under the worldsheet Weyl symmetry
$\gamma_{ij}\rightarrow\Lambda^2(z)\gamma_{ij}$. Requiring cancellation
of the anomalies in this symmetry at the quantum level gives
differential-equation restrictions on the background fields
($g_{\sst MN}$, $A_{\sst MN}$, $\phi$) that may be viewed as effective
equations of motion for these massless modes.\cite{cfmp} This system of
effective equations may be summarized by the corresponding field-theory
effective action
\bea
\lefteqn{I_{\rm eff}=\int
d^Dx\sqrt{-g}e^{-2\phi}\Big[(D-26)-\ft32\alpha'(R+4\nabla^2\phi
-4(\nabla\phi)^2}\hspace{4cm}\nonumber\\
&&-\ft1{12}F_{\sst MNP}F^{\sst MNP}+{\cal O}(\alpha')^2\Big]\ ,\label{ieff}
\eea
where
$F_{\sst MNP} =\partial_{\sst M}A_{\sst NP} +
\partial_{\sst N}A_{\sst PM} + \partial_{\sst P}A_{\sst MN}$ is the
3-form field strength for the $A_{\sst MN}$ gauge potential. The $(D-26)$
term reflects the critical dimension for the bosonic string: flat space is
a solution of the above effective theory only for
$D=26$.

     The effective action for superstring theories contains a similar
(NS-NS) sector, but with the substitution of $(D-26)$ by $(D-10)$,
reflecting the different critical dimension for superstrings. In
addition, superstring theories have a Ramond-Ramond (R-R) sector
of further bosonic fields. For example, the type IIA theory~~\cite{IIA} has
R-R field strengths $F_{[2]}=dA_{[1]}$ and
$F_{[4]}=dA_{[3]}+A_{[1]}\wedge F_{[3]}$, where the $[n]$
subscripts indicate the ranks of the forms. In the type IIB theory,\cite{IIB}
on the other hand, one has $F_{[1]}=d\chi$, where $\chi$ is a R-R
zero-form ({\it i.e.}\  a pseudoscalar field), $F^{\rm
R}_{[3]}=dA^{\rm R}_{[2]}$, a second 3-form field strength making a pair
together with $F^{\rm NS}_{[3]}$ from the NS-NS sector, and
$F_{[5]}=dA_{[4]}$, which is a {\em self-dual} 5-form in $D=10$,
$F_{[5]}={^\ast}F_{[5]}$.

     Thus one naturally encounters field strengths of ranks 1--5 in
type II theories. In addition, one may use $\epsilon_{[10]}$ to
dualize certain field strengths; {\it e.g.}\  the original $F_{[3]}$
may be dualized to ${^\ast}F_{[3]}$, which is a 7-form. The upshot is
that, in considering solutions to string-theory effective field
equations, antisymmetric-tensor gauge field strengths of diverse
ranks need to be taken into account. These field strengths will play
an essential r\^ole in supporting the $p$-brane solutions that we
shall describe.

     The effective action (\ref{ieff}) is written in the form
directly obtained from string $\sigma$-model calculations. It is not
written in the form generally preferred by relativists, which has a
clean Einstein-Hilbert term free from exponential prefactors like
$e^{-2\phi}$. One may rewrite the effective action in a different
{\em frame} by making a Weyl-rescaling field redefinition
$g_{\sst MN}\rightarrow e^{\lambda\phi}g_{\sst MN}$. $I_{\rm eff}$ as
written in (\ref{ieff}) is in the {\em string frame}; after an
integration by parts, it takes the form
\be
I^{\rm \,string}=\int d^{10}x\sqrt{-g^{{\rm (s)}}}e^{-2\phi}\Big[R(g^{{\rm
(s)}})+4\nabla_{\sst M}\phi\nabla^{\sst
M}\phi-\ft1{12}F_{\sst MNP}F^{\sst MNP}\Big]\ .\label{istring}
\ee
After making the transformation
$g^{\rm (e)}_{\sst MN}=e^{-\phi/2}g^{\rm (s)}_{\sst MN}$, one
obtains the {\em Einstein frame} action,
\be
I^{\rm \,Einstein}=\int d^{10}x\sqrt{-g^{\rm (e)}}\Big[R(g^{\rm (e)}) -
\ft12\nabla_{\sst M}
\phi\nabla^{\sst M}\phi-\ft1{12}e^{_\phi}F_{\sst MNP}F^{\sst MNP}\Big]\ ,
\label{ieinstein}
\ee
where the indices are now raised and lowered with $g^{\rm (e)}_{\sst
MN}$. To understand how this Weyl rescaling works, note that under
$x$-independent rescalings, the connection $\Gamma_{\sst MN}{}^{\sst
P}$ is invariant. This carries over also to terms with $\phi$
undifferentiated, which emerge from the $e^{\lambda\phi}$ Weyl
transformation. One then chooses $\lambda$ so as to eliminate the
$e^{-2\phi}$ factor. Terms with $\phi$ undifferentiated do change,
however. As one can see in (\ref{ieinstein}), the Weyl transformation is
just what is needed to unmask the positive-energy sign of the kinetic term 
for the $\phi$ field, despite the apparently negative sign of its kinetic
term in $I^{{\rm \,string}}$.

\section{The $p$-brane ansatz}\label{sec:pbraneans}
\subsection{General action and field equations}\label{ssec:genact}

     Motivated by the above summary of the effective
field theories derived from string theories, let us now consider a
classical system in $D$ dimensions comprising the metric
$g_{\sst MN}$, a scalar field $\phi$ and an $(n-1)$-form gauge potential
$A_{[n-1]}$ with corresponding field strength $F_{[n]}$, the
whole described by the action
\be
I=\int D^Dx\sqrt{-g}\Big[R-\ft12\nabla_{\sst M}\phi\nabla^{\sst M}\phi
-\ft{1}{2n!}e^{a\phi}F^2_{[n]}\Big]\ .\label{igen}
\ee
We shall consider later in more detail how (\ref{igen}) may be obtained by
a consistent truncation from a full supergravity theory in $D$ dimensions.
The value of the parameter $a$ controlling the interaction of the scalar
field $\phi$ with the field strength $F_{[n]}$ in (\ref{igen}) will vary
according to the cases considered in the following.

     Varying the action (\ref{igen}) produces the following
set of equations of motion:
\begin{subeqnarray}
R_{MN} &=& \ft12\partial_{\sst M}\phi\partial_{\sst N}\phi +
S_{\sst MN}\\
S_{MN} &=& {1\over
2(n-1)!}e^{a\phi}(F_{{\sst M}\cdots}F_{\sst N}{}^{\cdots} -{n-1\over
n(D-2)}F^2 g_{\sst MN})\\
\makebox[0pt]{\hspace{2.5cm}$\nabla_{{\sst M}_1}(e^{a\phi}F^{{\sst
M}_1\cdots {\sst M}_n}) ~~ = ~~ 0$}\\
\square\phi &=& {a\over 2n!}e^{a\phi}F^2\ .\label{eqmots}
\end{subeqnarray}

\subsection{Electric and magnetic ans\"atze}\label{ssec:ansatze}

     In order to solve the above equations, we shall make a simplifying
ansatz. We shall be looking for solutions preserving certain unbroken
supersymmetries, and these will in turn require unbroken
translational symmetries as well. For simplicity, we shall also require
isotropic symmetry in the directions ``transverse'' to the
translationally-symmetric ones. These restrictions can subsequently be
relaxed in generalizations of the basic class of $p$-brane solutions that
we shall discuss here. For this basic class of solutions, we make an
ansatz requiring $\hbox{ (Poincar\'e)}_d\times {\rm SO}(D-d)$ symmetry.
One may view the sought-for solutions as flat $d=p+1$ dimensional
hyperplanes embedded in the ambient $D$-dimensional spacetime; these
hyperplanes may in turn be viewed as the histories, or worldvolumes, of
$p$-dimensional spatial surfaces. Accordingly, let the spacetime
coordinates be split into two ranges: $x^{\sst M}=(x^\mu,y^m)$, where
$x^\mu$ ($\mu=0,1,\cdots,p=d-1$) are coordinates adapted to the
$\hbox{(Poincar\'e)}_d$ isometries on the worldvolume and where $y^m$
($m=d,\cdots,D-1$) are the coordinates ``transverse'' to the
worldvolume.

     An ansatz for the spacetime metric that respects the $\hbox{
(Poincar\'e)}_d\times {\rm SO}(D-d)$ symmetry is~~\cite{dghrr}
\be
ds^2 = e^{2A(r)}dx^\mu dx^\nu\eta_{\mu\nu} +
e^{2B(r)}dy^mdy^n\delta_{mn}\ ,\label{ansatz}
\ee
where $r=\sqrt{y^my^m}$ is the isotropic radial coordinate in the
transverse space. Since the metric components depend only on $r$,
translational invariance in the worldvolume directions $x^\mu$ and ${\rm
SO}(D-d)$ symmetry in the transverse directions $y^m$ is guaranteed.

     The corresponding ansatz for the scalar field $\phi(x^M)$ is
simply $\phi=\phi(r)$.

     For the antisymmetric tensor gauge field, we face a
bifurcation of possibilities for the ansatz, the two possibilities
being related by duality. The first possibility is naturally expressed
directly in terms of the gauge potential $A_{[n-1]}$. Just as the
Maxwell 1-form naturally couples to the worldline of a charged
particle, so does $A_{[n-1]}$ naturally couple to the worldvolume of
a $p=d-1=(n-1)-1$ dimensional ``charged'' extended object. The
``charge'' here will be obtained from Gauss'-law surface
integrals involving $F_{[n]}$, as we shall see later. Thus, the first
possibility for $A_{[n-1]}$ is to support a $d_{\rm el}=n-1$
dimensional worldvolume. This is what we shall call the ``elementary,'' or
``electric'' ansatz:
\be
A_{\mu_1\cdots\mu_{n-1}} = \epsilon_{\mu_1\cdots\mu_{n-1}}e^{C(r)}
\ ,\hspace{.5cm}\mbox{others zero.}\label{elans}
\ee
${\rm SO}(D-d)$ isotropicity and $\hbox{(Poincar\'e)}_d$
symmetry are guaranteed here because the function $C(r)$ depends only on
the transverse radial coordinate $r$. Instead of the ansatz (\ref{elans}),
expressed in terms of
$A_{[n-1]}$, we could equivalently have given just the $F_{[n]}$ field
strength:
\be
F^{\rm(el)}_{m\mu_1\cdots\mu_{n-1}} =
\epsilon_{\mu_1\cdots\mu_{n-1}}\partial_me^{C(r)}\ ,
\hspace{.5cm}\mbox{others zero.}\label{felans}
\ee
The worldvolume dimension for the elementary ansatz
(\ref{elans},\ref{felans}) is clearly $d_{\rm el}=n-1$.

     The second possible way to relate the rank $n$ of $F_{[n]}$ to the
worldvolume dimension $d$ of an extended object is suggested by considering
the dualized field strength ${^\ast}F$, which is a $(D-n)$ form. If one
were to find an underlying gauge potential for ${^\ast}F$ (locally
possible by courtesy of a Bianchi identity), this would naturally couple
to a $d_{\rm so}=D-n-1$ dimensional worldvolume. Since such a dualized
potential would be nonlocally related to the fields appearing in the
action (\ref{igen}), we shall not explicitly follow this
construction, but shall instead take this reference to the
dualized theory as an easy way to identify the worldvolume dimension for
the second type of ansatz. This ``solitonic'' or ``magnetic'' ansatz for
the antisymmetric tensor field is most conveniently expressed in terms
of the field strength $F_{[n]}$, which now has nonvanishing values only
for indices corresponding to the transverse directions:
\be
F^{\rm (mag)}_{m_1\cdots m_n} = \lambda\epsilon_{m_1\cdots
m_np}{y^p\over r^{n+1}}\ ,\hspace{.5cm}
\mbox{others zero,}\label{magans}
\ee
where the magnetic-charge parameter $\lambda$ is a constant of
integration, the only thing left undetermined by this ansatz. The
power of $r$ in the solitonic/mag\-netic ansatz is determined by requiring
$F_{[n]}$ to satisfy the Bianchi identity.\footnote{Specifically, one
finds $\partial_qF_{m_1\cdots m_n} = r^{-(n+1)}\big(\epsilon_{m_1\cdots
m_nq}-(n+1)\epsilon_{m_1\cdots m_np}y^py_q/r^2\big)$; upon taking the
totally antisymmetrized combination $[qm_1\cdots m_n]$, the factor of
$(n+1)$ is evened out between the two terms and then one finds from
cycling a factor $\sum_my^my_m=r^2$, thus obtaining cancellation.} Note
that the worldvolume dimensions of the elementary and solitonic cases are
related by $d_{\rm so}= \tilde d_{\rm el}\equiv D-d_{\rm el}-2$; note
also that this relation is idempotent, {\it i.e.}\ $\widetilde{(\tilde
d)}=d$.

\subsection{Curvature components and $p$-brane
equations}\label{ssec:pbraneqs}

In order to write out the field equations after insertion of the
above ans\"atze, one needs to compute the Ricci tensor for the
metric.\cite{stainless} This is most easily done by introducing veilbeins,
{\it i.e.,} orthonormal frames,\cite{mtw} with tangent-space indices
denoted by underlined indices:
\be
g_{\sst MN} = e_{\sst M}{}^{\sst
\underline E}e_{\sst N}{}^{\sst\underline F}
\eta_{\sst\underline E\,\underline F}\ .\label{vielbs}
\ee
Next, one constructs the corresponding 1-forms:
$e^{\sst\underline E}=dx^{\sst M}e_{\sst M}{}^{\sst\underline E}$.
Splitting up the tangent-space indices
${\st\underline E}=(\underline\mu,\underline m)$ similarly to the
world indices ${\st M}=(\mu,m)$, we have for our ans\"atze the veilbein
1-forms
\be
e^{\underline\mu}=e^{A(r)}dx^\mu\ ,\hspace{2cm}
e^{\underline m}=e^{B(r)}dy^m\ .\label{1forms}
\ee

     The corresponding spin connection 1-forms are determined by the
condition that the torsion vanishes,
$de^{\sst\underline
E}+\omega^{\sst\underline E}{}_{\sst\underline F}\wedge
e^{\sst\underline F}=0$, which yields
\bea
\omega^{\underline\mu\,\underline\nu}&=&0\ ,\hspace{2cm}
\omega^{\underline\mu\,\underline n}
=e^{-B(r)}\partial_nA(r)e^{\underline\mu}\nonumber\\
\omega^{\underline m\,\underline n}
&=& e^{-B(r)}\partial_nB(r)e^{\underline m}\ 
-\ e^{-B(r)}\partial_mB(r)e^{\underline n}\ .\label{conn1forms}
\eea
The curvature 2-forms are then given by
\be
R_{[2]}^{\sst\underline E\,\underline F} =
d\omega^{\sst\underline E\,\underline F} +
\omega^{\sst\underline E\,\underline D}\wedge
\omega_{\sst\underline D}{}^{\sst\underline F}\ .\label{curv2f}
\ee
{}From the curvature components so obtained, one finds the Ricci tensor
components
\bea
R_{\mu\nu} &=& -\eta_{\mu\nu}e^{2(A-B)}(A''+d(A')^2+\tilde dA'B' +
{(\tilde d+1)\over r}A')\nonumber\\
R_{mn} &=& -\delta_{mn}(B''+dA'B'+\tilde d(B')^2+{(2\tilde d+1)\over
r}B' + {d\over r}A')\\
&&-{y^my^n\over r^2}(\tilde
B''+dA''-2dA'B'+d(A')^2-\tilde d(B')^2-{\tilde d\over r}B'-{d\over
r}A')\ ,\nonumber\label{riccicomps}
\eea
where again, $\tilde d=D-d-2$, and the primes indicate
$\partial/\partial r$ derivatives.

     Substituting the above relations, one finds the set of equations
that we need to solve to obtain the metric and $\phi$:
\be
\begin{array}{rclr}
A''+d(A')^2+\tilde dA'B'+{(\tilde d+1)\over r}A' &\makebox[0pt]{=}&
{\tilde d\over 2(D-2)}S^2\hspace{1cm} &\{\mu\nu\}\\
B''+dA'B'+\tilde d(B')^2+{(2\tilde d+1)\over r}B'+{d\over r}A'
&\makebox[0pt][1]{=}& -{d\over 2(D-2)}S^2&\{\delta_{mn}\}\\
\tilde B'' + dA''-2dA'B'+d(A')^2-\tilde d(B')^2\hspace{1cm}\\
-{\tilde d\over r}B'-{d\over r}A'+\ft12(\phi')^2 &\makebox[0pt]{=}&
\ft12S^2 &\{y_my_n\}\\
\phi''+dA'\phi'+\tilde dB'\phi'+{(\tilde d+1)\over r}\phi'
&\makebox[0pt]{=}& -\ft12\varsigma aS^2\ ,
\end{array}\label{pbraneqs}
\ee
where $\varsigma=\pm 1$ for the elementary/solitonic cases and the source
appearing on the RHS of these equations is
\be
S = \left\{\begin{array}{ll}
(e^{\fft12 a\phi-dA+C})C'\hspace{1cm} &\mbox{elementary: $d=n-1$,
$\varsigma=+1$}\\
\lambda(e^{\fft12 a\phi-\tilde dB})r^{-\tilde d-1}&\mbox{solitonic:
$d=D-n-1$, $\varsigma=-1$.}
\end{array}\right.\label{pbranesource}
\ee

\subsection{$p$-brane solutions}\label{ssec:pbranesols}

     The $p$-brane equations (\ref{pbraneqs},\ref{pbranesource}) are
still rather daunting. In order to proceed further, we are going to
take a hint from the requirements for supersymmetry preservation, which
shall be justified in more detail later on. Accordingly, we shall now
refine our ans\"atze by imposing the linearity condition
\be
dA'+\tilde d B' = 0\ .\label{dadtb}
\ee
After eliminating $B$ using (\ref{dadtb}), the independent equations
become~~\cite{dilatonic}
\begin{subeqnarray}
\nabla^2\phi &=& -\ft12\varsigma aS^2\label{symeqsa}\\
\nabla^2A &=& {\tilde d\over 2(D-2)}S^2\label{symeqsb}\\
d(D-2)(A')^2 + \ft12\tilde d(\phi')^2 &=& \ft12\tilde dS^2
\ ,\label{symeqsc}
\end{subeqnarray}
where, for spherically-symmetric ({\it i.e.}\ isotropic) functions
in the transverse $(D-d)$ dimensions, the Laplacian is $\nabla^2\phi
= \phi'' + (\tilde d+1)r^{-1}\phi'$. 

     Equations (\ref{symeqsa}a,b) suggest that we now further refine
the ans\'atze by imposing another linearity condition:
\be
\phi' = {-\varsigma a(D-2)\over\tilde d}A'\ .\label{phirel}
\ee
At this stage, it is useful to introduce a new piece of notation,
letting
\be
a^2 = \Delta-{2d\tilde d\over(D-2)}\ .\label{Delta}
\ee
With this notation, equation (\ref{symeqsc}c) gives
\be
S^2 = {\Delta(\phi')^2\over a^2}\ ,\label{s2}
\ee
so that the remaining equation for $\phi$ becomes $\nabla^2\phi +
{\varsigma\Delta\over2a}(\phi')^2 = 0$, which can be re-expressed as a
Laplace equation,\footnote{Note that Eq.\ (\ref{laplace}) can also be
more generally derived; for example, it still holds if one relaxes the
assumption of isotropicity in the transverse space.}
\be
\nabla^2e^{{\varsigma\Delta\over2a}\phi} = 0\ .\label{laplace}
\ee
Solving this in the transverse $(D-d)$ dimensions with our assumption of
transverse isotropicity ({\it i.e.}\ spherical symmetry) yields
\be
e^{{\varsigma\Delta\over 2a}\phi} = H(y) = 1 + {k\over r^{\tilde d}}
\hspace{1.5cm}k>0\ ,\label{phisol}
\ee
where the constant of integration $\phi\for_{r\rightarrow\infty}$ has
been set equal to zero here for simplicity: $\phi_\infty = 0$. The
integration constant $k$ in (\ref{phisol}) sets the mass scale of the
solution; it has been taken to be positive in order to ensure the absence
of naked singularities at finite $r$. This positivity restriction is
similar to the usual restriction to a positive mass parameter $M$ in the
standard Schwarzshild solution.

     In the case of the elementary/electric ansatz, with
$\varsigma=+1$, it still remains to find the function $C(r)$ that
determines the antisymmetric-tensor gauge field potential. In this
case, it follows from (\ref{pbranesource}) that
$S^2=e^{a\phi-1dA}(C'e^C)^2$. Combining this with (\ref{symeqsc}), one
finds the relation
\be
{\partial\over\partial r}(e^C)={-\sqrt\Delta\over a}e^{-\fft12 a\phi
+ dA}\phi'\label{crel}
\ee
(where it should be remembered that $a<0$). Finally, it is
straightforward to verify that the relation (\ref{crel}) is consistent 
with the equation of motion for $F_{[n]}$:
\be
\nabla^2 C + C'(C' + \tilde B' - dA' + a\phi') = 0\ .\label{ceqn}
\ee

     In order to simplify the explicit form of the solution, we now
pick values of the integration constants to make
$A_\infty=B_\infty=0$, so that the solution tends to flat empty space
at transverse infinity. Assembling the result, starting from the
Laplace-equation solution $H(y)$ (\ref{phisol}), one
finds~~\cite{dkl,stainless}
\begin{subeqnarray}
ds^2 &=& H^{-4\tilde d\over\Delta(D-2)}dx^\mu dx^\nu\eta_{\mu\nu} +
H^{4d\over\Delta(D-2)}dy^mdy^m\\
e^\phi &=&
H^{2a\over\varsigma\Delta}\hspace{1cm}\varsigma
=\cases{+1,&elementary/electric\cr-1,&solitonic/magnetic\cr}\\
H(y) &=& 1+{k\over r^{\tilde d}}\ ,\label{pbranesol}
\end{subeqnarray}
and in the elementary/electric case, $C(r)$ is given by
\be
e^C = {2\over\sqrt\Delta}H^{-1}\ .\label{csol}
\ee
In the solitonic/magnetic case, the constant of integration is related
to the magnetic charge parameter $\lambda$ in the ansatz (\ref{magans})
by
\be
k = {\sqrt\Delta\over2\tilde d}\lambda\ .\label{klambdarel}
\ee
In the elementary/electric case, this relation may be taken to {\em
define} the parameter $\lambda$.

\section{Examples}\label{sec:examples}

     Consider now the bosonic sector of $D=11$ supergravity, which
has the action
\be
I_{11} = \int d^{11}x\left\{\sqrt{-g}(R-\ft1{48}F_{[4]}^2) + \ft16
F_{[4]}\wedge F_{[4]}\wedge A_{[3]}\right\}\ .\label{D11act}
\ee
There are two particular points to note about this action.
The first is that no scalar field is present. This follows from the
supermultiplet structure of the $D=11$ theory, in which all fields
are gauge fields. In lower dimensions, of course, scalars do appear;
{\it e.g.}\ the dilaton in $D=10$ type IIA supergravity emerges out of
the $D=11$ metric upon dimensional reduction from $D=11$ to $D=10$.
The absence of the scalar that we had in our general discussion may
be handled here simply by identifying the scalar coupling parameter $a$
with zero, so that the scalar may be consistently truncated from our
general action (\ref{igen}). Since $a^2=\Delta-2d\tilde d/(D-2)$, we
identify $\Delta=2\cdot3\cdot6/9=4$ for the $D=11$ cases.

The second point to note is the presence of the $FFA$ Chern-Simons
term in (\ref{D11act}). This term is required by $D=11$ local
supersymmetry, with the coefficient as given in (\ref{D11act}). Under
the bosonic antisymmetric-tensor gauge transformation $\delta
A_{[3]}=d\Lambda_{[2]}$, the $FFA$ term in (\ref{D11act}) is invariant
(up to a total derivative) separately from the kinetic term. 

     In our general discussion given above in Sec.\ \ref{sec:pbraneans},
we did not take into account the effects of such $FFA$ terms. This
omission, however, is not essential to the basic class of $p$-brane
solutions that we are studying. Note that for $n=4$, the $F_{[4]}$
antisymmetric tensor field strength supports either an
elementary/electric solution with $d=n-1=3$ ({\it i.e.}\ a $p=2$
membrane) or a solitonic/magnetic solution with $\tilde d=11-3-2=6$
({\it i.e.}\  a $p=5$ brane). In both these elementary and solitonic
cases, the $FFA$ term in the action (\ref{D11act}) vanishes and hence
this term does not make any non-vanishing contribution to the metric
field equations for our ans\"atze. For the antisymmetric tensor field
equation, a further check is necessary, since there one requires
the {\it variation} of the $FFA$ term to vanish in order to consistently
ignore it. The field equation for $A_{[3]}$ is
\be
\partial_{\sst M}\left(\sqrt{-g}F^{\sst MUVW}\right) +
{1\over2(4!)^2}\epsilon^{{\sst
UVW}x_1x_2x_3x_4y_1y_2y_3y_4}F_{x_1x_2x_3x_4}F_{y_1y_2y_3y_4}=0\
.\label{Feq}
\ee 
By direct inspection, one sees that the second term in this equation
vanishes for both ans\"atze.

Next, we shall consider the elementary/electric and the solitonic/magnetic
$D=11$ cases in detail. Subsequently, we shall explore how these
particular solutions fit into wider, ``black,'' families of $p$-branes.

\subsection{$D=11$ Elementary/electric 2-brane}\label{ssec:electric}

     From our general discussion in Sec.\ \ref{sec:pbraneans}, we have
the elementary-ansatz solution~~\cite{ds}
\be\begin{array}{rcl}
ds^2 &\makebox[0pt]{=}& (1+{k\over r^6})^{-\sffrac23}d^\mu
dx^\nu\eta_{\mu\nu} + (1+{k\over
r^6})^{-\sffrac13}dy^mdy^m\\
A_{\mu\nu\lambda} &\makebox[0pt]{=}& \epsilon_{\mu\nu\lambda}
(1+{k\over r^6})^{-1}\ ,\hspace{.75cm}
\mbox{other components zero.}\\
&&\multicolumn{1}{r}{\mbox{\{\underline{electric 2-brane: isotropic
coordinates}\}}}\end{array}\label{isoel2br}
\ee
At first glance, this solution looks like it might be singular at
$r=0$. However, if one calculates the invariant components of the
curvature tensor $R_{\sst MNPQ}$ and of the field strength
$F_{m\mu_1\mu_2\mu_3}$, subsequently referred to an orthonormal frame by
introducing vielbeins as in (\ref{1forms}), one finds these
invariants to be nonsingular. Moreover, although the proper distance
to the surface $r=0$ along a $t=x^0=\mbox{const.}$ geodesic diverges,
the surface $r=0$ can be reached along null geodesics in finite
affine parameter.\cite{dgt}

     Thus, one may suspect that the metric as given in
(\ref{isoel2br}) does not in fact cover the entire spacetime, and so one
should look for an analytic extension of it. Accordingly, one may consider
a change to ``Schwarzshild-type'' coordinates by setting $r=(\tilde
r^6-k)^{\sffrac16}$. The solution then becomes:\,\cite{dgt}
\be\begin{array}{rclr}
ds^2 &\makebox[0pt]{=}& (1+{k\over r^6})^{\sffrac23}(-dt^2 + d\sigma^2
+d\rho^2) + (1+{k\over r^6})^{-2}d\tilde r^2 + \tilde
r^2d\Omega_7^2\\ 
A_{\mu\nu\lambda} &\makebox[0pt]{=}&
\epsilon_{\mu\nu\lambda} (1+{k\over r^6})\ ,
&\llap{other components zero,}\\
&&\multicolumn{2}{r}{\mbox{\{\underline{electric 2-brane:
Schwarzshild-type coordinates}\}}}
\end{array}\label{shwel2br}
\ee
where we have supplied explicit worldvolume coordinates
$x^\mu=(t,\sigma,\rho)$ and where $d\Omega_7^2$ is the line element on
the unit 7-sphere, corresponding to the boundary $\partial{\cal
M}_{8\rm T}$ of the $11-3=8$ dimensional transverse space.

     The Schwarzshild-like coordinates make the surface $\tilde
r=k^{\sffrac16}$ (corresponding to $r=0$) look like a horizon. One may
indeed verify that the normal to this surface is a {\em null} vector,
confirming that $\tilde r=k^{\sffrac16}$ is in fact a horizon. This
horizon is {\em degenerate}, however. Owing to the \ffrac23 exponent
in the $g_{00}$ component, curves along the $t$ axis for $\tilde
r<k^{\sffrac16}$ remain timelike, so that light cones do not ``flip
over'' inside the horizon, unlike the situation for the classic
Schwarzshild solution.

     In order to see the structure of the membrane spacetime more
clearly, let us change coordinates once again, setting $\tilde
r=k^{\sffrac16}(1-R^3)^{-\sffrac16}$. Overall, the transformation from
the original isotropic coordinates to these new ones is effected by
setting $\tilde r=k^{\sffrac16}R^{\sffrac12}/(1-R^3)^{\sffrac16}$. In these
new coordinates, the solution becomes~~\cite{dgt}
\be\begin{array}{rclr}
ds^2 &\makebox[0pt]{=}& \left\{R^2(-dt^2 + d\sigma^2 + d\rho^2) +
4k^{\sffrac13}R^{-2}dR^2\right\} +
k^{\sffrac13}d\Omega_7^2&\hspace{1cm}\hfill\mbox{(a)}\\
&&+ k^{\sffrac13}[(1-R^3)^{-\sffrac13}-1][4R^{-2}dR^2 +
d\Omega_7^2]&\hfill\mbox{(b)}\\
A_{\mu\nu\lambda} &\makebox[0pt]{=}&
R^3\epsilon_{\mu\nu\lambda}\ ,
\hfill\mbox{other components zero.}\\
&&\multicolumn{1}{r}{\mbox{\{\underline{electric 2-brane: interpolating
coordinates}\}}}
\end{array}\label{intel2br}
\ee

     This form of the solution makes it clearer that the light-cones
do not ``flip over'' in the region inside the horizon (which is now at
$R=0$, with $R<0$ being the interior). The main usefulness of the third
form (\ref{intel2br}) of the membrane solution, however, is that it
reveals how the solution {\em interpolates} between other ``vacuum''
solutions of $D=11$ supergravity.\cite{dgt} As $R\rightarrow1$, the
solution becomes flat, in the asymptotic exterior transverse region.
As one approaches the horizon at $R=0$, line (b) of the metric in
(\ref{intel2br}) vanishes at least linearly in $R$. The residual
metric, given in line (a), may then be recognized as a standard form of
the metric on $(\mbox{AdS})_4\times{\cal S}^7$, generalizing the
Robinson-Bertotti solution on  $(\mbox{AdS})_2\times{\cal S}^2$ in
$D=4$. Thus, the membrane solution interpolates between flat space as
$R\rightarrow1$ and  $(\mbox{AdS})_4\times{\cal S}^7$ as
$R\rightarrow0$ at the horizon.

     Continuing on inside the horizon, one eventually encounters a
true singularity at $\tilde r=0$ ($R\rightarrow-\infty$). Unlike the
singularity in the classic Schwarzshild solution, which is spacelike
and hence unavoidable, the singularity in the membrane spacetime is
{\em timelike.} Generically, geodesics do not intersect the
singularity at a finite value of an affine parameter value. Radial null
geodesics do intersect the singularity at finite affine parameter,
however, so the spacetime is in fact genuinely singular. The timelike
nature of this singularity, however, invites one to consider coupling a
$\delta$-function {\em source} to the solution at $\tilde r=0$. Indeed,
the $D=11$ supermembrane action,\cite{bst} which generalizes the
Nambu-Goto action for the string, is the unique  ``matter'' system that
can consistently couple to $D=11$ supergravity.\cite{bst,dhis} Analysis
of this coupling yields a relation between the parameter $k$ in the
solution (\ref{isoel2br}) and the {\em tension} $T$ of the supermembrane
action:\,\cite{ds}
\be
k={\kappa^2T\over3\Omega_7}\ ,\label{ktens}
\ee
where $1/(2\kappa^2)$ is the coefficient of $\sqrt{-g}R$ in the
Einstein-Hilbert Lagrangian and $\Omega_7$ is the volume of the unit
7-sphere ${\cal S}^7$, {\it i.e.}\ the solid angle subtended by the
boundary at transverse infinity.

\begin{figure}
\epsfbox{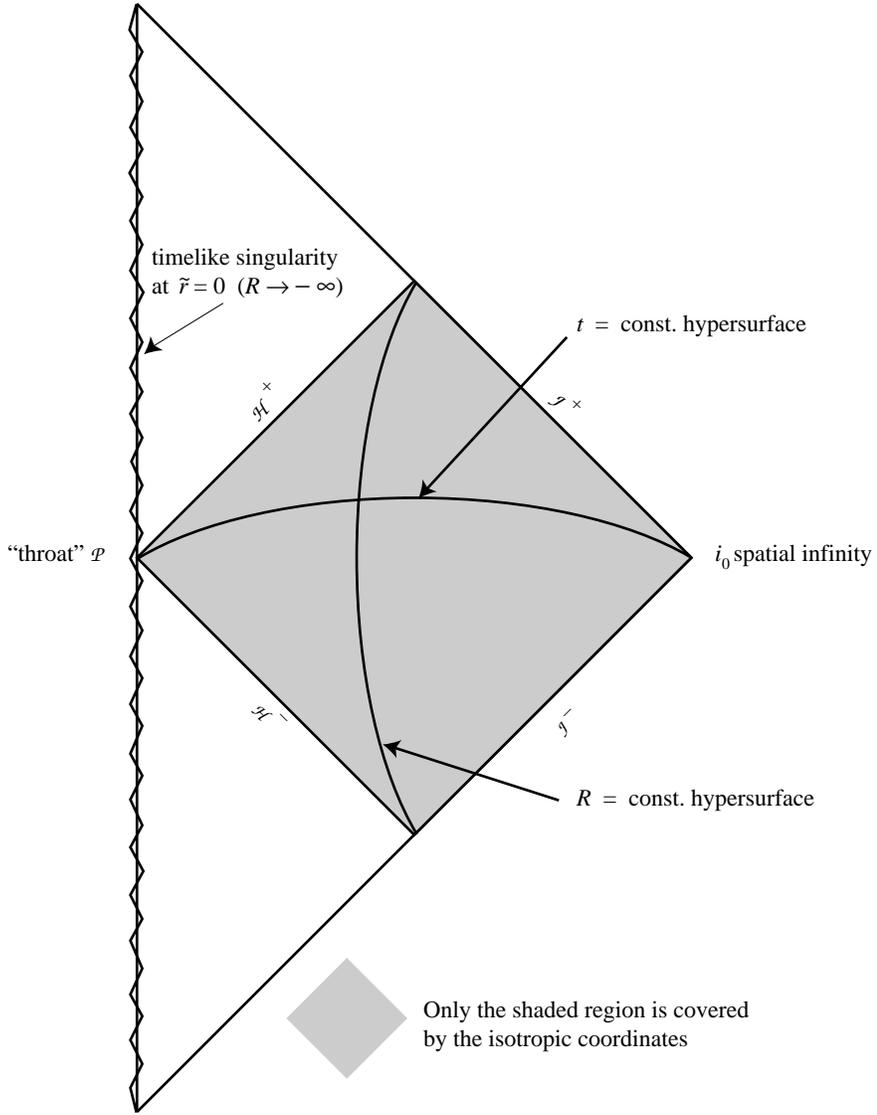}
\caption{Carter-Penrose diagram for the $D=11$ elementary/electric 2-brane
solution.\label{fig:D11elcp}}
\end{figure}

     The global structure of the membrane spacetime~~\cite{dgt} is
similar to the extreme Reissner-Nordstrom solution of General
Relativity.\cite{hawkel} This global structure is summarized by a
Carter-Penrose diagram as shown in Figure \ref{fig:D11elcp}, in which
the angular coordinates on ${\cal S}^7$ and also two ignorable
worldsheet coordinates have been suppressed. As one can see, the region
mapped by the isotropic coordinates does not cover the whole spacetime.
This region, shaded in the diagram, is geodesically incomplete, since
one may reach its boundaries ${\cal H}^+$, ${\cal H}^-$ along radial
null geodesics at a finite affine-parameter value. These boundary
surfaces are not singular, but, instead, constitute future and past
horizons (one can see from the form (\ref{shwel2br}) of the solution
that the normals to these surfaces are null). The ``throat'' $\cal P$ in
the diagram should be thought of as an exceptional point at infinity,
and not as a part of the central singularity.

\begin{figure}[h]
\epsfbox{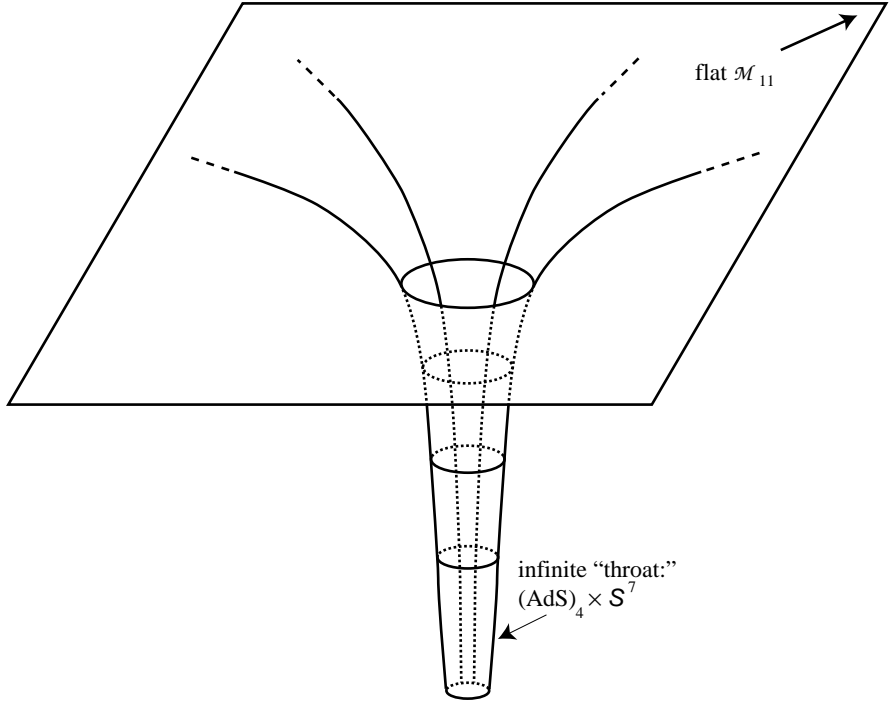}
\caption{The $D=11$ elementary/electric 2-brane solution interpolates
between flat space at ${\cal J}^\pm$ and $(\mbox{AdS})_4\times{\cal
S}^7$ at the horizon.\label{fig:funnel}}
\end{figure}

     The region exterior to the horizon interpolates between flat
regions ${\cal J}^\pm$ at future and past null infinities and a
geometry that asymptotically tends to $(\mbox{AdS})_4\times{\cal
S}^7$ on the horizon. This interpolating portion of the spacetime,
corresponding to the shaded region of Figure \ref{fig:D11elcp}
which is covered by the isotropic coordinates, may be sketched as shown
in Figure \ref{fig:funnel}.

\clearpage

\subsection{$D=11$ Solitonic/magnetic 5-brane}\label{ssec:magnetic}

     Now consider the 5-brane solution to the $D=11$ theory given by
the solitonic ansatz for $F_{[4]}$. In isotropic
coordinates, this solution is a magnetic 5-brane:\,\cite{guv}
\crampest\be\begin{array}{rclr}
ds^2 &\makebox[0pt]{=}& (1+{k\over r^3})^{-\sffrac13}dx^\mu
dx^\nu\eta_{\mu\nu} + (1+{k\over
r^3})^{\sffrac23}dy^mdy^m\hspace{.6cm}&\mu,\nu=0,\cdots,5\\ F_{m_1\cdots
m_4} &\makebox[0pt]{=}& 3k\epsilon_{m_1\cdots m_4 p}{y^p\over r^5}
&\llap{other components zero.}\\
&&&\llap{\{\underline{magnetic 5-brane: isotropic coordinates}\}}
\end{array}\label{mag5br}
\ee\uncramp

     As in the case of the elementary/electric membrane, this solution
interpolates between two ``vacua'' of $D=11$ supergravity. Now,
however, these asymptotic geometries consist of the flat region
encountered as $r\rightarrow\infty$ and of $(\mbox{AdS})_7\times{\cal
S}^4$ as one approaches $r=0$, which once again is a degenerate
horizon. Combining two coordinate changes analogous to those of
the elementary case, $r=(\tilde r^3-k)^{\sffrac13}$ and $\tilde r =
k^{\sffrac13}(1-R^6)^{-\sffrac13}$, one has an overall transformation
\be
r = {k^{\sffrac13}R^2\over(1-R^6)^{\sffrac13}}\ .\label{coordtransf}
\ee
After these coordinate changes, the metric becomes
\be\begin{array}{rcl}
ds^2 &\makebox[0pt]{=}& R^2dx^\mu dx^\nu\eta_{\mu\nu} +
k^{\sffrac23}\left[{4R^{-2}(1+R^6)^2\over(1-R^6)^{\sffrac83}}dR^2 +
{d\Omega_4^2\over(1-R^6)^{\sffrac23}}\right]\ .\\
&&\multicolumn{1}{r}{\mbox{\{\underline{magnetic 5-brane: interpolating
coordinates}\}\hspace{.3cm}}}
\end{array}\label{symmag5br}
\ee

     Once again, the surface $r=0 \leftrightarrow R=0$ may be seen from 
(\ref{symmag5br}) to be a nonsingular degenerate horizon. In this
case, however, not only do the light cones maintain their timelike
orientation when crossing the horizon, as already happened in the
electric case (\ref{intel2br}), but now the magnetic solution
(\ref{symmag5br}) is in fact fully {\em symmetric}~~\cite{ght} under a
discrete isometry $R\rightarrow -R$.

\begin{figure}
\epsfbox{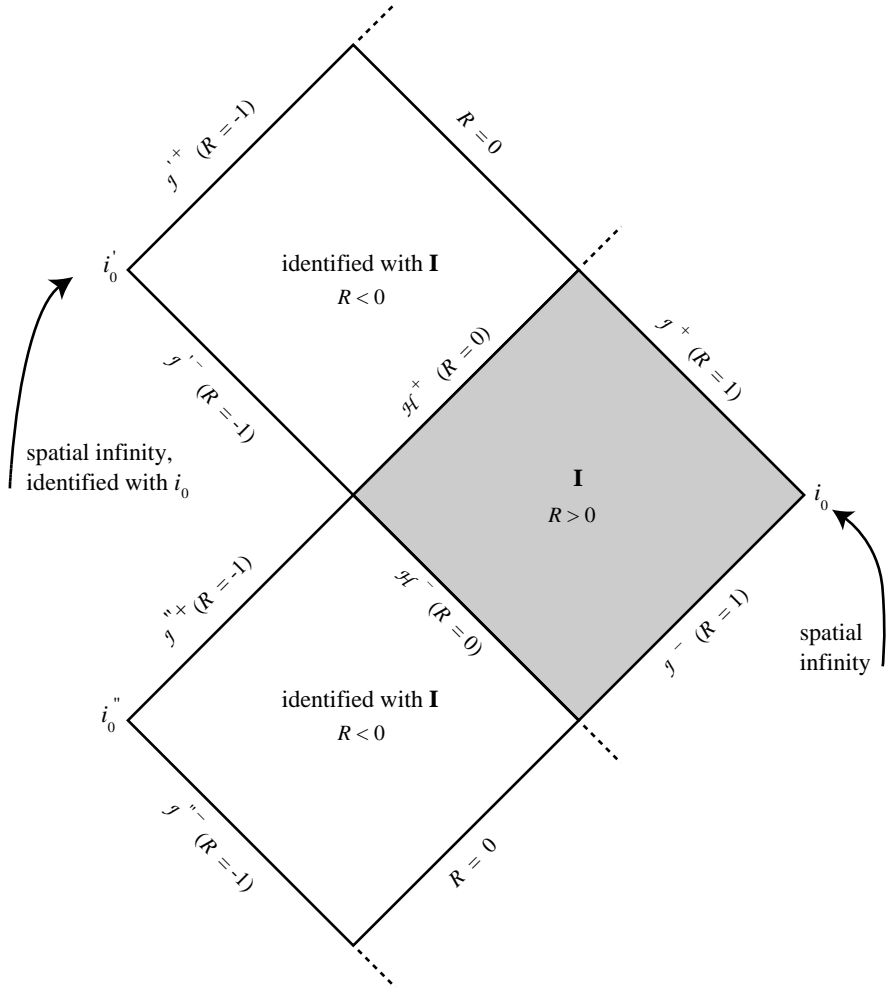}
\caption{Carter-Penrose diagram for the solitonic/magnetic
5-brane solution.\label{fig:D11cpmag}}
\end{figure}

     Given this isometry $R\rightarrow -R$, one can {\em identify}
the spacetime region $R\le 0$ with the region $R\ge 0$. This
identification is analogous to the identification one naturally makes
for flat space when written in polar coordinates, with the metric
$ds^2_{\mbox{flat}}=-dt^2+dr^2+r^2d^2$. Ostensibly, in these
coordinates there appear to be separate regions of flat space with
$r\gtlt 0$, but, owing to the existence of the isometry $r\rightarrow
-r$, these regions may be identified. Accordingly, in the
solitonic/magnetic 5-brane spacetime, we identify the region $-1<R\le
0$ with the region $0\le R<1$. In the asymptotic limit where
$R\rightarrow -1$, one finds                           an asymptotically
flat geometry that is indistinguishable from the region where
$R\rightarrow +1$, {\it i.e.}\ where $r\rightarrow\infty$. Thus, there
is no singularity at all in the solitonic/magnetic 5-brane geometry.
There is still an infinite ``throat,'' however, at the horizon, and the
region covered by the isotropic coordinates might again be sketched as
in Figure \ref{fig:funnel}, except now with the asymptotic geometry down
the ``throat'' being $(\mbox{AdS})_7\times{\cal S}^4$ instead of
$(\mbox{AdS})_4\times{\cal S}^7$ as for the elementary/electric
solution. The Carter-Penrose diagram for the solitonic/magnetic
5-brane solution is given in Figure \ref{fig:D11cpmag}, where the full
diagram extends indefinitely by ``tiling'' the section shown. Upon
using the $R\rightarrow -R$ isometry to make discrete
identifications, however, the whole of the spacetime may be
considered to consist of just region {\bf I}, which is the region
covered by the isotropic coordinates (\ref{mag5br}).

     After identification of the $R\gtlt 0$ regions, the
5-brane spacetime (\ref{mag5br}) is {\em geodesically complete.}
Unlike the case of the elementary membrane solution
(\ref{isoel2br},\ref{intel2br}), one finds in the solitonic/magnetic
case that the null geodesics passing through the horizon at $R = 0$
continue to evolve in their affine parameters without bound as
$R\rightarrow -1$. Thus, the solitonic 5-brane solution is {\em
completely non-singular.}

\clearpage

     The electric and magnetic $D=11$ solutions discussed here and in
the previous subsection are special in that they do not involve a
scalar field, since the bosonic sector of $D=11$ supergravity
(\ref{D11act}) does not even contain a scalar field. Similar solutions
occur in other situations where the parameter $a$ (\ref{Delta}) for a
field strength supporting a $p$-brane solution vanishes, in which cases
the scalar fields may consistently be set to zero; this happens for
$(D,d)=(11,3)$, (11,5), (10,4), (6,2), (5,1), (5,2) and (4,1). In these
special cases, the solutions are nonsingular at the horizon and so one
may analytically continue through to the other side of the horizon. When
$d$ is even for ``scalarless'' solutions of this type, there exists a
discrete isometry analogous to the $R\rightarrow -R$ isometry of the
$D=11$ 5-brane solution (\ref{symmag5br}), allowing the outer and inner
regions to be identified.\cite{ght} When $d$ is odd in such cases, the
analytically-extended metric eventually reaches a timelike curvature
singularity at $\tilde r=0$.

     When $a\ne0$ and the scalar field associated to the field strength
supporting a solution cannot be consistently set to zero, then the
solution is singular at the horizon, as can be seen directly in the
scalar solution (\ref{phisol}) itself (where we recall that in
isotropic coordinates, the horizon occurs at $r=0$)

\subsection{Black branes}\label{ssec:blackbranes}

     In order to understand better the family of
supergravity solutions that we have been discussing, let us now
consider a generalization that lifts the degenerate nature of the
horizon. Written in Schwarzshild-type coordinates, one finds the
generalized ``black brane'' solution~~\cite{hsdl,dlp}
\be\begin{array}{rcl}
ds^2 &\makebox[0pt]{=}& -{\Sigma_+\over\Sigma_-^{\left\{1-
{4\tilde d\over\Delta(D-2)}\right\}}}dt^2 + \Sigma_-^{{4\tilde
d\over\Delta(D-2)}}dx^idx^i\\
&&\hspace{.5cm} + {\Sigma_-^{\left\{{2a^2\over\Delta\tilde
d}-1\right\}}\over\Sigma_+}d\tilde r^2 + \tilde
r^2\Sigma_-^{2a^2\over\Delta\tilde d}d\Omega^2_{D-d-1}\\
e^{{\varsigma\Delta\over2a}\phi} &\makebox[0pt]{=}& \Sigma_-^{-1}
\hspace{2cm}\Sigma_\pm = 1-\left(r_\pm\over\tilde r\right)^{\tilde
d}\ .\\
&&\multicolumn{1}{r}{\mbox{\{\underline{black brane:
Schwarzshild-type coordinates}\}}}
\end{array}\label{blackbrane}
\ee
The antisymmetric tensor field strength for this solution
corresponds to a charge parameter $\lambda = 2\tilde
d/\sqrt\Delta(r_=r_-)^{\tilde d/2}$, either electric or magnetic.

     The characteristic feature of the above ``blackened'' $p$-branes is
that they have a nondegenerate, nonsingular outer horizon at $\tilde
r=r_+$, at which the light cones ``flip over.'' At $\tilde r =r_-$, one
encounters an inner horizon, which, however, coincides in general with a
curvature singularity. The singular nature of the solution at $\tilde r
= r_-$ is apparent in the scalar $\phi$ in (\ref{blackbrane}). For
solutions with $p\ge1$, the singularity at the inner horizon persists
even in cases where the scalar $\phi$ is absent. 

     The {\em extremal} limit of the black brane solution occurs for
$r_+=r_-$. When $a=0$ and scalars may consistently be set to zero, the
singularity at the horizon $r_+=r_-$ disappears and then one may
analytically continue through the horizon. In this case, the light cones
do not ``flip over'' at the horizon because one is really crossing two
coalesced horizons, and the coincident ``flips'' of the light cones cancel
out.

     The generally singular nature of the inner horizon of the non-extreme
solution (\ref{blackbrane}) shows that the ``location'' of the $p$-brane
in spacetime should normally be thought to coincide with the inner
horizon, or with the degenerate horizon in the extremal case.

\section{Masses, Charges and Supersymmetry}\label{sec:masschargesusy}

     The $p$-brane solutions that we have been studying are supported by
antisymmetric tensor gauge field strengths that fall off at transverse
infinity like $r^{-(\tilde d+1)}$, as one can see from
(\ref{felans},\ref{csol},\ref{magans}). This asymptotic falloff is slow
enough to give a nonvanishing total charge density from a Gauss' law flux
integral at transverse infinity, and we shall see that, for the
``extremal'' class of solutions that is our main focus, the mass density
of the solution saturates a ``Bogomol'ny bound'' with respect to the
charge density. This relation between densities is in turn connected to
another feature of these solutions: although purely bosonic, they
preserve unbroken some portion of the original supersymmetry of the
corresponding supergravity theory.\cite{dghrr}

\subsection{Masses}\label{ssect:masses}

     Let us begin with the mass density. Since the $p$-brane solutions
have translational symmetry in their $p$ spatial worldvolume directions,
the total energy as measured by a surface integral at spatial infinity
diverges, owing to the infinite extent. What is thus more appropriate to
consider instead is the value of the density, energy/(unit $p$-volume).
Since we are considering solutions in their rest frames, this will also
give the value of mass/(unit $p$-volume), or {\em tension} of the
solution. Instead of the standard spatial $d^{D-2}\Sigma^a$ surface
integral, this will be a $d^{(D-d-1)}\Sigma^m$ surface integral over the
boundary $\partial{\cal M}_{\rm\sst T}$ of the transverse space.

     The ADM formula for the energy density written as a Gauss'-law
integral (see, {\it e.g.,} Ref.\cite{mtw}) is, dropping the
divergent spatial $d\Sigma^{\mu=i}$ integral,
\be
{\cal E} = {1\over 4\Omega_{D-d-1}}\int_{\partial{\cal M}_{\rm
T}}d^{D-d-1}\Sigma^m(\partial^nh_{mn}-\partial_mh^b_b)\ ,\label{admenerg}
\ee
written for $g_{\sst MN} = \eta_{\sst MN} + h_{\sst MN}$ tending
asymptotically to flat space in Cartesian coordinates, and with $a,b$
{\em spatial} indices running over the values $\mu=i=1,\ldots,d-1$;
$m=d,\ldots,D-1$. $\Omega_{D-d-1}$ is the volume of the
${\cal S}^{D-d-1}$ unit sphere. For the general $p$-brane solution
(\ref{pbranesol}), one finds
\be
h_{mn}={4kd\over\Delta(D-2)r^{\tilde d}}\delta_{mn}\ ,\hspace{2cm} h^b_b
= {8k(d+\ft12 \tilde d)\over\Delta(D-2)r^{\tilde d}}\ ,\label{hasymp}
\ee
and, since $d^{(D-d-1)}\Sigma^m = r^{\tilde d}y^md\Omega^{(D-d-1)}$, one
finds
\be
{\cal E} = {k\tilde d\over\Delta}\ ,\label{energy}
\ee
and, recalling that $k=\sqrt\Delta\lambda/(2\tilde d)$, we
consequently have a relation between the mass per unit $p$ volume and the
charge parameter of the solution
\be
{\cal E} = {\lambda\over2\sqrt\Delta}\ .\label{masschargerel}
\ee

     By contrast, the black brane solution (\ref{blackbrane}) has ${\cal
E}>\lambda/(2\sqrt\Delta)$, so the extremal $p$-brane solution
(\ref{pbranesol}) is seen to {\em saturate} the inequality ${\cal E}\ge
\lambda/(2\sqrt\Delta)$.

\subsection{Charges}\label{ssect:charges}

     As one can see from (\ref{energy},\ref{masschargerel}), the relation
(\ref{klambdarel}) between the integration constant $k$ in the solution
(\ref{pbranesol}) and the charge parameter $\lambda$ implies a deep link
between the energy density and certain electric or magnetic
charges. In the electric case, this charge is a quantity conserved by
virtue of the equations of motion for the antisymmetric tensor gauge
field $A_{[n-1]}$, and has generally become known as a ``Page charge,''
after its first discussion in Ref.\cite{dp} To be specific, if we once
again consider the bosonic sector of $D=11$ supergravity theory
(\ref{D11act}), for which the antisymmetric tensor field equation was
given in (\ref{Feq}), one finds the Gauss'-law form conserved
quantity~~\cite{dp} 
\be
U={1\over4\Omega_7}\int_{\partial{\cal M}_8}({^\ast}F_{[4]} + \ft12
A_{[3]}\wedge F_{[4]})\ ,\hspace{2cm}\mbox{\{\underline{electric
charge}\}}
\label{pagecharge}
\ee
where the integral of the 7-form integrand is over the boundary at
infinity of an arbitrary infinite 8-dimensional spacelike
subspace of $D=11$ spacetime. This arbitrariness of choice
in the 8-dimensional spacelike subspace means that (\ref{pagecharge}) 
in fact represents a whole set of conserved charges. A basis of these
charges may be obtained by taking the embedding of ${\cal M}_8$ into
the 10-dimensional spatial hypersurface to be specified by a
volume-element 2-form. Accordingly, the electric Page charge
(\ref{pagecharge}) should properly be denoted by $U_{\sst AB}$.

     For the $p$-brane solutions (\ref{pbranesol}), the $\int A\wedge F$
term in (\ref{pagecharge}) vanishes. The $\int{^\ast}F$ term does,
however, give a contribution in the elementary/electric case,
provided one picks ${\cal M}_8$ to be the transverse space to the $d=3$
membrane worldvolume, ${\cal M}_{8{\rm T}}$. The surface element for this
transverse space is $d\Sigma^m_{(7)}$, so for the $p=2$ elementary
membrane solution (\ref{isoel2br}), one finds
\be
U={1\over4\Omega_7}\int_{\partial M_{8{\rm T}}}d\Sigma^m_{(7)}F_{m012} =
{\lambda\over4}\ .\label{membcharge}
\ee
Since the $D=11$ $F_{[4]}$ field strength supporting this solution has
$\Delta=4$, the mass/charge relation is
\be
{\cal E} = U = {\lambda\over4}\ .\label{elmasschargerel}
\ee
Thus, like the classic extreme Reissner-Nordstrom black-hole solution
to which it is strongly related (as can be seen from the Carter-Penrose
diagram given in Figure \ref{fig:D11elcp}), the $D=11$ membrane solution
has equal mass and charge densities, saturating the inequality ${\cal
E}\ge U$.

     Given the existence of an electric-type charge (\ref{pagecharge}),
one also expects to find a magnetic-type charge, which, however, should
be conserved {\em topologically,} {\it i.e.}\  by virtue of the Bianchi
identity $dF_{[4]}=0$. This magnetic-type charge, being an integral over
a four-form $F_{[4]}$, necessarily again involves integration over a
submanifold of the spatial hypersurface of $D=11$ spacetime:
\be
V = {1\over4\Omega_4}\int_{\partial{\cal M}_5}F_{[4]}\ ;\hspace{2cm}
\mbox{\{\underline{magnetic charge}\}}\label{magpagecharge}
\ee
the surface integral now being taken over the boundary at infinity of a
spacelike 5-dimensional subspace. As with the electric-type Page charge
$U_{AB}$, $V$ really represents a whole multiplet of charges, depending
on the embedding of the subsurface ${\cal M}_5$ into the 10-dimensional
spatial hypersurface. This embedding may be specified in terms a volume
5-form, so the magnetic charge should properly be denoted by a 5-form
$V_{\sst ABCDE}$.

     It is the magnetic form of charge (\ref{magpagecharge}) that is
carried by the solitonic/mag\-netic 5-brane solution (\ref{mag5br}). Once
again, there is only one orientation of the subsurface ${\cal M}_5$ that
gives a nonvanishing contribution, {\it i.e.}\ that with ${\cal
M}_5={\cal M}_{5{\rm T}}$, the transverse space to the $d=6$ worldvolume:
\be
V = {1\over4\Omega_4}\int_{\partial{\cal M}_{5{\rm
T}}}d\Sigma^m_{(4)}\epsilon_{mnpqr}F^{npqr} = {\lambda\over4}\
.\label{5brcharge}
\ee
Thus, in the solitonic/magnetic 5-brane case as well, we have a
saturation of the mass-charge inequality:
\be
{\cal E} = V = {\lambda\over4}\ .\label{magmasschargerel}
\ee

\subsection{Supersymmetry}\label{ssect:supersymmetry}

     Since the bosonic solutions that we have been considering are {\em
consistent truncations} of $D=11$ supergravity, they must also possess
another conserved quantity, the {\em supercharge.} Admittedly, since the
supercharge is a Grassmanian (anticommuting) quantity, its value will
clearly be zero for the class of purely bosonic solutions that we have
been discussing. However, the functional form of the supercharge is still
important, as it determines the form of the asymptotic supersymmetry
algebra. The Gauss'-law form of the supercharge is given as
an integral over the boundary of the spatial hypersurface. For the $D=11$
solutions, this surface of integration is the boundary at
infinity $\partial{\cal M}_{10}$ of the $D=10$ spatial hypersurface; the
supercharge is then~~\cite{cjs}
\be
Q = \int_{\partial{\cal M}_{10}}\Gamma^{0bc}\psi_cd\Sigma_{(9)b}\ .
\label{supercharge}
\ee
One can also rewrite this in fully Lorentz-covariant form, where
$d\Sigma_{(9)b} = d\Sigma_{(9)0b}\rightarrow d\Sigma_{(9){\sst AB}}$:
\be
Q = \int_{\partial{\cal M}_{10}}\Gamma^{\sst ABC}\psi_{\sst
C}d\Sigma_{(9){\sst AB}}\ .\label{lorentzsupercharge}
\ee

     After appropriate definitions of Poisson brackets, the
$D=11$ supersymmetry algebra for the supercharge
(\ref{supercharge},\ref{lorentzsupercharge}) is found to be~~\cite{agit}
\be
\{Q,Q\} = C(\Gamma^{\sst A}P_{\sst A} + \Gamma^{\sst AB}U_{\sst AB} +
\Gamma^{\sst ABCDE}V_{\sst ABCDE})\ ,\label{susyalg}
\ee
where $C$ is the charge conjugation matrix, $P_{\sst A}$ is the
energy-momentum 11-vector and $U_{\sst AB}$ and $V_{\sst ABCDE}$ are
electric- and magnetic-type charges of precisely the sorts discussed in
the previous subsection. Thus, the supersymmetry algebra wraps together
all of the conserved Gauss'-law type quantities that we have discussed.

     The positivity of the $Q^2$ operator on the LHS of the algebra
(\ref{susyalg}) is at the root of the {\em Bogomol'ny
bounds}~~\cite{gh,ght,lp}
\begin{subeqnarray}
{\cal E} &\ge& (2/\sqrt\Delta)U\hspace{2cm}
\mbox{\{\underline{electric}\}}\\
{\cal E} &\ge& (2/\sqrt\Delta)V\hspace{2cm}
\mbox{\{\underline{magnetic}\}}\label{bogbounds}
\end{subeqnarray}
that are saturated by the $p$-brane solutions.

     The saturation of the Bogomol'ny inequalities by the $p$-brane
solutions is an indication that they fit into special types of
supermultiplets. All of these bound-saturating solutions share the
important property that they {\em leave some portion of the
supersymmetry unbroken}. Within the family of $p$-brane solutions that
we have been discussing, it turns out~~\cite{lp} that the $\Delta$ values
of such ``supersymmetric'' $p$-branes are of the form $\Delta=4/N$, where
$N$ is the number of antisymmetric tensor field strengths participating
in the solution (distinct, but of the same rank). The different charge
contributions to the supersymmetry algebra occurring for different
values of $N$ (hence different $\Delta$) affect the Bogomol'ny bounds as
shown in (\ref{bogbounds}).

     In order to see how a purely bosonic solution may leave some
portion of the supersymmetry unbroken, consider specifically again the
membrane solution of $D=11$ supergravity.\cite{ds} This theory~~\cite{cjs}
has just one spinor field, the gravitino $\psi_{\sst M}$. Checking for
the consistency of setting $\psi_{\sst M}=0$ with the supposition of some
residual supersymmetry with parameter
$\epsilon(x)$ requires solving the equation
\be
\delta\psi_{\sst A}\for_{\psi=0} = \tilde D_{\sst A}\epsilon = 0\
,\label{deltapsi}
\ee
where $\psi_{\sst A}=e_{\sst A}{}^{\sst M}\psi_{\sst M}$ and
\bea
\tilde D_{\sst A}\epsilon &=& D_{\sst A}\epsilon -
{1\over288}\left(\Gamma_{\sst A}{}^{\sst BCDE} - 8\delta_{\sst
A}{}^{\sst B}\Gamma^{\sst CDE}\right)F_{\sst BCDE}\epsilon\nonumber\\
D_{\sst A}\epsilon &=& (\partial_{\sst A} + \ft14 \omega_{\sst
A}{}^{\sst BC}\Gamma_{\sst BC})\epsilon\ .\label{dtildeps}
\eea
Solving the equation $\tilde D_{\sst A}\epsilon = 0$ amounts to finding
a {\em Killing spinor} field in the presence of the bosonic background.
Since the Killing spinor equation (\ref{deltapsi}) is linear in
$\epsilon(x)$, the Grassmanian (anticommuting) character of this
parameter is irrelevant to the problem at hand, which thus reduces
effectively to solving (\ref{deltapsi}) for a commuting quantity.

     In order to solve the Killing spinor equation (\ref{deltapsi}) in a
$p$-brane background, it is convenient to adopt an appropriate basis for
the $D=11$ $\Gamma$ matrices. For the $d=3$ membrane background, one
would like to preserve ${\rm SO}(2,1)\times {\rm SO}(8)$ covariance. An
appropriate basis that does this is
\be
\Gamma_{\sst A} =
(\gamma_\mu\otimes\Sigma_9,\oneone_{(2)}\otimes\Sigma_m)\
,\label{gammasplit}
\ee
where $\gamma_\mu$ and $\oneone_{(2)}$ are $2\times2$ ${\rm SO}(2,1)$
matrices; $\Sigma_9$ and $\Sigma_m$ are $16\times16$ ${\rm SO}(8)$
matrices, with $\Sigma_9 = \Sigma_3\Sigma_4\ldots\Sigma_{10}$, so
$\Sigma_9^2 = \oneone_{(16)}$. The most general spinor field consistent
with $\hbox{(Poincar\'e)}_3\times {\rm SO}(8)$ invariance in this spinor
basis is of the form
\be
\epsilon(x,y) =\epsilon_2\otimes\eta(r)\ ,\label{epsilonform}
\ee
where $\epsilon_2$ is a {\em constant} ${\rm SO}(2,1)$ spinor and
$\eta(r)$ is an ${\rm SO}(8)$ spinor depending only on the isotropic
radial coordinate $r$; $\eta$ may be further decomposed into $\Sigma_9$
eigenstates by the use of $\ft12(\oneone\pm\Sigma_9)$ projectors.

     Analysis of the the Killing spinor condition (\ref{deltapsi}) in
the above spinor basis leads to the following requirements on the
background and on the spinor field $\eta(r)$:\,\cite{dghrr,ds}
\begin{enumerate}
\item The background must satisfy the conditions $3A' + 6B' = 0$ and
$C'e^C = 3A'e^{3A}$. The first of these conditions is, however, precisely
the linearity-condition refinement (\ref{dadtb}) that we made in the
$p$-brane ansatz; the second condition follows from the ansatz refinement
(\ref{phirel}) (considered as a condition on $\phi'/a$) and from
(\ref{crel}). Thus, what appeared previously to be simplifying
specializations in the derivation given in Section \ref{sec:pbraneans}
turn out in fact to be conditions {\em required} for supersymmetric
solutions.
\item $\eta(r) = e^{-C(r)/6}\eta_0$, where $\eta_0$ is a constant ${\rm
SO}(8)$ spinor. Note that, after imposing this requirement, at most a
finite number of parameters can remain unfixed in the product spinor
$\epsilon\eta_0$; {\it i.e.}\ the {\em local} supersymmetry of the $D=11$
theory is almost entirely broken by any particular solution. The
maximum number of {\em rigid} unbroken supersymmetry components is
achieved for $D=11$ flat space, which has a full 32-component rigid
supersymmetry.
\item $(\oneone-\Sigma_9)\eta_0 = 0$, so the constant ${\rm SO}(8)$
spinor $\eta_0$ is also required to be {\em chiral}.\footnote{The
specific chirality indicated here is correlated with the sign choice
made in the elementary/electric form ansatz (\ref{elans}); one may
accordingly observe from (\ref{D11act}) that a $D=11$  parity
transformation requires a sign flip of $A_{[3]}$.} This cuts the number
of surviving parameters in the product $\epsilon\eta_0$ by {\em half:}
the total number of surviving rigid supersymmetries in $\epsilon(x,y)$
is thus $2\cdot 8 =16$ (real spinor components). Since this is half
of the maximum possible number ({\it i.e.}\  half of that for flat
space), one says that the membrane solution ``preserves half'' of the
supersymmetry.
\end{enumerate}

     Similar consideration of the solitonic/magnetic 5-brane
solution~~\cite{guv} (\ref{mag5br}) shows that it also preserves half the
supersymmetry in the above sense. Half preservation is the maximum that
can be achieved short of an empty-space solution, and when this happens,
it corresponds to the existence of zero eigenvalues of the operator
$\{Q,Q\}$. The positive semi-definiteness of this operator is the
underlying principle in the derivation of the Bogomol'ny bounds
(\ref{bogbounds}).\cite{gh,ght,lp} A consequence of this
positive semi-definiteness is that zero eigenvalues correspond to
solutions that saturate the Bogomol'ny inequalities
(\ref{bogbounds}), and these solutions preserve one
component of unbroken supersymmetry for each such zero eigenvalue.

\section{Kaluza-Klein dimensional reduction}\label{sec:kkred}

     Let us return now to the arena of purely bosonic field theories,
and consider the relations between various bosonic-sector theories and the
corresponding relations between $p$-brane solutions. It is
well-known that supergravity theories are related by dimensional
reduction from a set of basic theories, the largest of which being
$D=11$ supergravity. The spinor sectors of the theories are equally well
related by dimensional reduction, but in the following, we shall
restrict our attention to the purely bosonic sector.

     In order to set up the procedure, let us consider a theory in
$(D+1)$ dimensions, but break up the metric in $D$-dimensionally
covariant pieces:
\be
d\hat s^2 = e^{2\alpha\varphi}ds^2 + e^{2\beta\phi}(dz + {\cal A}_{\sst
M}dx^{\sst M})^2\label{kkmetricans}
\ee
where carets denote $(D+1)$-dimensional quantities corresponding to the
$(D+1)$-dimensional coordinates $x^{\hat{\sst M}}=(x^{\sst M},z)$;
$ds^2$ is the line element in $D$ dimensions and $\alpha$ and $\beta$
are constants. The $D$-scalar $\varphi$ emerges from $(D+1)$ dimensions
as $(2\beta)^{-1}\ln g_{zz}$. Adjustment of the constants $\alpha$ and
$\beta$ is necessary to obtain desired structures in $D$ dimensions. In
particular, one should pick $\beta=-(D-2)\alpha$ in order to have the
Einstein-frame form of the gravitational action in $(D+1)$ dimensions go
over to the Einstein-frame form of the action in $D$ dimensions.

     The essential step in a Kaluza-Klein dimensional reduction is a
{\em consistent truncation} of the field variables, generally made
by choosing them to be independent of the reduction coordinate $z$. By
consistent truncation, we always mean a restriction on the variables that
commutes with variation of the action to produce the field equations, {\it
i.e.}\ a restriction such that solutions to the equations for the
restricted variables are also solutions to the equations for the
unrestricted variables. This ensures that the lower-dimensional solutions
that we shall obtain are also particular solutions to higher-dimensional
supergravity equations as well. Making the parameter choice
$\beta=-(D-2)\alpha$ to preserve the Einstein-frame form of the action,
one obtains
\be
\sqrt{-\hat g}R\Big(\hat g) = \sqrt{-g}(R(g) -
(D-1)(D-2)\alpha^2\nabla_{\sst M}\varphi\nabla^{\sst M}\varphi -
\ft14e^{-2(D-1)\alpha\varphi}{\cal F}_{\sst MN}{\cal F}^{\sst MN}\Big)
\label{actionred}
\ee
where ${\cal F}=d{\cal A}$. If one now chooses $\alpha^2 =
[2(D-1)(D-2)]^{-1}$, the $\varphi$ kinetic term becomes
conventionally normalized.

     Next, one needs to establish the reduction ansatz for the
$(D+1)$-dimen\-sional antisymmetric tensor gauge field $\hat F_{[n]} =
d\hat A_{[n-1]}$. Clearly, among the $n-1$ antisymmetrized indices of
$\hat A_{[n-1]}$ at most one can take the value $z$, so we have the
decomposition
\be
\hat A_{[n-1]} = B_{[n-1]} + B_{[n-2]}\wedge dz\ .\label{kkantisymtens}
\ee
All of these reduced fields are to be taken to be functionally
independent of $z$. For the corresponding field strengths, first define
\begin{subeqnarray}
G_{[n]} &=& dB_{[n-1]}\\
G_{[n-1]} &=& dB_{[n-2]}\ .\label{redfstrengths}
\end{subeqnarray}
However, these are not exactly the most convenient quantities to
work with, since a certain ``Chern-Simons'' structure appears upon
dimensional reduction. The metric in $(D+1)$ dimensions couples to all
fields, and, consequently, dimensional reduction will produce some terms
with undifferentiated Kaluza-Klein vector fields ${\cal A}_{\sst M}$
coupling to $D$-dimensional antisymmetric tensors. Accordingly, it
is useful to introduce
\be
G'_{[n]} = G_{[n]}-G_{[n-1]}\wedge{\cal A}\ ,\label{kkchsimG}
\ee
where the second term in (\ref{kkchsimG}) may be viewed as a
Chern-Simons correction from the reduced $D$-dimensional point of view.

     At this stage, we are ready to perform the dimensional reduction of
our general action (\ref{igen}). We find
\be
\hat I =: \int d^{D+1}x\sqrt{-\hat g}\Big[R(\hat g) - \ft12
\nabla_{\hat{\sst M}}\phi\nabla^{\hat{\sst M}}\phi - {1\over2n!}e^{\hat
a\phi}\hat F^2_{n]}\Big]\label{Ihat}
\ee
reduces to
\bea
I &=& \int d^Dx\sqrt{-g}\Big[R - \ft12\nabla_{\sst M}\phi\nabla^{\sst
M}\phi - \ft12\nabla_{\sst M}\varphi\nabla^{\sst M}\varphi -
\ft14e^{-2(D-1)\alpha\varphi}{\cal F}^2_{[2]}\nonumber\\
&&~~-{1\over2n!}e^{-2(n-1)\alpha\varphi + \hat\alpha\phi}G'^2_{[n]} -
{1\over2(n-1)!}e^{2(D-n)\alpha\varphi+\hat a\phi}G^2_{[n-1]}\Big]\
.\label{Ired}
\eea
Although the dimensional reduction (\ref{Ired}) has
produced a somewhat complicated result, the important point to note is
that each of the $D$-dimensional antisymmetric-tensor field strength
terms $G'^2_{[n]}$ and $G^2_{[n-1]}$ has an exponential prefactor of the
form $e^{a_r\tilde\phi_r}$, where the $\tilde\phi_r$, $r=(n,n-1)$ are
$SO(2)$-rotated combinations of $\varphi$ and $\phi$. Now, keeping just one
while setting to zero the other two of the three gauge fields $({\cal
A}_{[1]}, B_{[n-2]}, B_{n-1]})$, but retaining at the same time the
scalar-field combination appearing in the corresponding exponential
prefactor, is a {\em consistent truncation.} Thus, any one of the three
field strengths $({\cal F}_{[2]}, G_{[n-1]}, G'_{[n]})$, retained alone
together with its corresponding scalar-field combination, can support
$p$-brane solutions in $D$ dimensions of the form that we have been
discussing.

     An important point to note here is that in each of the
$e^{a_r\tilde\phi}$ prefactors, the coefficient $a_r$ satisfies
\be
a_r^2 = \Delta - {2d_r\tilde d_r\over(D-2)} = \Delta -
{2(r-1)(D-r-1)\over(D-2)}\label{deltared}
\ee
with the {\em same} value of $\Delta$ as for the ``parent'' coupling
parameter $\hat a$, satisfying
\be
\hat a_r^2 = \Delta - {2d_{(n)}\tilde d_{(n)}\over((D+1)-2)} = \Delta -
{2(n-1)(D-n)\over(D-1}\label{deltahigh}
\ee
in $D+1$ dimensions. Thus, although the individual parameters $a_r$ are
both $D$- and $r$-dependent, the quantity $\Delta$ is {\em preserved} under
Kaluza-Klein reduction for both of the ``descendant'' field-strength
couplings (to $G'^2_{[n]}$ or to $G^2_{[n-1]}$) coming from the
original term $e^{\hat a\phi}\hat F_{[n]}^2$. The 2-form field strength
${\cal F}_{[2]} = d{\cal A}$, on the other hand, emerges out of the
gravitational action in $D+1$ dimensions; its coupling parameter
corresponds to $\Delta=4$.

     If one retains in the reduced theory only one of the field
strengths (${\cal F}_{[2]}$, $G_{[n-1]}$, $G'_{[n]}$), together with its
corresponding scalar-field combination, then one finds oneself back in the
situation described by our general action (\ref{igen}), and then the $p$
brane solutions obtained for the general case in Sec.\ \ref{sec:pbraneans}
immediately become applicable. Moreover, since retaining only one (field
strength, scalar) combination in this way effects a consistent truncation
of the theory, solutions to this simple truncated system are {\em also}
solutions to the untruncated theory, and indeed are also solutions to
the original $(D+1)$-dimensional theory, since the Kaluza-klein
dimensional reduction is also a consistent truncation.

     The $p$-brane solutions are ideally structured for Kaluza-Klein
reduction, because they are independent of the ``worldvolume'' $x^\mu$
coordinates. Accordingly, one may let the reduction coordinate $z$ be one
of the $x^\mu$. Consequently, the only thing that needs to be done
to such a solution in order to {\em reinterpret} it as a solution of the
reduced system (\ref{Ired}) is to perform a Weyl rescaling on it in order
to be in accordance with the form given in the Kaluza-Klein ansatz, which
was adjusted so as to maintain the Einstein-frame form of the gravitational
term in the action.

     After making such a reinterpretation, elementary/solitonic $p$-branes
in $(D+1)$ dimensions give rise to elementary/solitonic $(p-1)$-branes in
$D$ dimensions, corresponding to the {\em same} value of $\Delta$, as one
can see from (\ref{deltared},\ref{deltahigh}). Note that in this process,
the quantity $\tilde d$ is conserved, since both $D$ and $d$ reduce by one.
Reinterpretation of $p$ brane solutions in this way, corresponding to
standard Kaluza-Klein reduction on a worldvolume coordinate, proceeds 
diagonally on a $D$ versus $d$ plot, and hence is referred to as {\em
diagonal} dimensional reduction. This procedure is the analogue, for
supergravity field-theory solutions, of the procedure of {\em double
dimensional reduction}~~\cite{dhis} for $p$-brane worldvolume actions,
which can be taken to constitute the $\delta$-function {\em sources} for
singular $p$-brane solutions, coupled in to resolve the singularities.

\section{Multiple field-strength solutions}\label{sec:multifsols}

     Upon Kaluza-Klein dimensional reduction by repeated single steps down
to $D$ dimensions, the bosonic sector of maximal supergravity
(\ref{D11act}) reduces to~~\cite{lp}
\bea
I_D &=& \int d^Dx\sqrt{-g}\Big[R-\ft12(\partial\vec \phi)^2 -
\ft1{48}e^{\vec a\cdot\vec\phi}F_{[4]}^2 - \ft1{12}\sum_ie^{\vec
a_i\cdot\vec\phi}(F^i_{[3]})^2\nonumber\\
&&\hspace{1.5cm}-\ft14\sum_{i<j}e^{\vec
a_{ij}\cdot\vec\phi}(F^{ij}_{[2]})^2 -
\ft14\sum_ie^{\vec b_i\cdot\phi}({\cal F}^i_{[2]})^2\label{I_D}\\
&&\hspace{1.5cm}-\ft12\sum_{i<j<k}e^{\vec
a_{ijk}\cdot\vec\phi}(F^{ijk}_{[1]})^2 -
\ft12\sum_{ij}e^{\vec b_{ij}\cdot\phi}({\cal F}^{ij}_{[1]})^2\Big] + {\cal
L}_{FFA}\ ,\nonumber
\eea
where $i,j = 1,\ldots,11-D$, and field strengths with multiple $i,j$
indices may be taken to be antisymmetric in those indices since these
``internal'' indices arise in the stepwise reduction procedure, and two
equal index values never occur in a multi-index sum. The
``straight-backed'' field strengths $F_{[4]}$, $F^i_{[3]}$, $F^{ij}_{[2]}$
and $F^{ijk}_{[1]}$ are descendants from $F_{[4]}$ in $D=11$. The
``calligraphic'' field strengths ${\cal F}^i_{[2]}$ are the field
strengths for Kaluza-Klein vectors like ${\cal A}_{\sst M}$ in
(\ref{kkmetricans}) that emerge from the higher-dimensional metric upon
dimensional reduction. Once such a Kaluza-Klein vector has appeared,
subsequent dimensional reduction gives rise also to 1-form field strengths 
${\cal F}^{ij}_{[1]}$ for zero-form gauge potentials ${\cal A}_{[0]}^{ij}$
as a consequence of the usual one-step reduction (\ref{kkantisymtens}) of a
1-form gauge potential.

     The scalar fields $\vec\phi$ that appear in the exponential
prefactors in (\ref{I_D}) form an $(11-D)$-vector of fields that may be
called ``dilatonic'' scalars. For each field strength occurring in
(\ref{I_D}), there is a corresponding ``dilaton vector'' of coefficients
determining the linear combination of the dilatonic scalars appearing in
its exponential prefactor. For the 4-, 3-, 2- and 1-form
``straight-backed'' field strengths emerging from $F_{[4]}$ in $D=11$,
these coefficients are denoted correspondingly $\vec a$, $\vec a_i$, $\vec
a_{ij}$ and $\vec a_{ijk}$; for the ``calligraphic'' field strengths
corresponding to Kaluza-Klein vectors and zero-form gauge potentials
emerging out of the metric, these are denoted $\vec b_i$ and $\vec b_{ij}$
correspondingly. Thankfully, not all of these dilaton vectors are
independent, and in fact, they may all be expressed in terms of the 4-form
and 3-form dilaton vectors $\vec a$ and
$\vec a_{ij}$:\,\cite{lp}
\be\begin{array}{rclrcl}
\vec a_{ij} &\makebox[0pt]{=}& \vec a_i&\hspace{1cm} b_i
&\makebox[0pt]{=}& -\vec a_i + \vec a\\
\vec a_{ijk}&\makebox[0pt]{=}&\vec a_i + \vec a_j +\vec a_k -
2\vec a&\hspace{1cm}\vec b_{ij}&\makebox[0pt]{=}&-\vec a_i + \vec a_j
\ .
\end{array}\label{dilvecrels}
\ee
Another important feature of the dilaton vectors is that they satisfy the
following dot product relations:
\bea
\vec a\cdot\vec a &=& {2(11-D)\over D-2}\nonumber\\
\vec a\cdot\vec a_i &=& {2(8-D)\over D-2}\\
\vec a_i\cdot\vec a_j &=& 2\delta_{ij} + {2(6-D)\over D-2}\ .\nonumber
\label{dilvecprods}
\eea

     Throughout this review, we have emphasized {\em consistent
truncations} in making simplifying restrictions of complicated systems of
equations, so that the solutions of a simplified system are nonetheless
perfectly valid solutions of the more complicated untruncated system. Once
again, with the equations of motion following from (\ref{I_D}) we face a
complicated system that calls for analysis in simplified subsectors, so
that we may use the solutions already found in our general study of the
action (\ref{igen}). Accordingly, we now seek a consistent truncation to
include just one dilatonic scalar combination
$\phi$ and one rank-$n$ field strength combination $F_{[n]}$, constructed
out of a certain number $N$ of ``retained'' field strengths
$F_{\alpha\,[n]}$, $\alpha=1,\ldots,N$, (possibly a
straight-backed/calligraphic mixture) selected from those appearing in
(\ref{I_D}), all the rest being set to zero.\cite{lp} Thus, we let
\be
\vec\phi=\vec n\phi + \vec\phi_\perp\ ,\label{phidecomp}
\ee
where $\vec n\cdot\vec\phi_\perp = 0$; in the truncation we seek to set
consistently $\vec\phi_\perp = 0$.

     Consistency for the field strengths $F_{\alpha\,[n]}$
requires them to be proportional.\cite{lp} We shall let the
dot product matrix for the retained field strengths be denoted
$M_{\alpha\beta}=:\vec a_\alpha\cdot \vec a_\beta$. Consistency of the
truncation requires that the
$\phi_\perp$ field equation be satisfied:
\be
\square\vec\phi_\perp - \sum_\alpha \Pi_\perp\cdot\vec
a_\alpha(F_{\alpha\,[n]})^2 = 0\ ,\label{phiperpeq}
\ee
where $\Pi_\perp$ is the projector into the dilaton-vector subspace
orthogonal to the retained dilaton direction $\vec n$. Setting
$\vec\phi_\perp = 0$ in (\ref{phiperpeq}) and letting the retained
$F_{\alpha\,[n]}$ be proportional, one sees that achieving consistency is
hopeless unless all the $e^{\vec a_\alpha\cdot\vec\phi}$ prefactors are
the same, requiring
\be
\vec a_\alpha\cdot\vec n = a \hspace{.5cm}\forall\alpha=1,\ldots,N\ ,
\label{anprod}
\ee
where the constant $a$ will play the role of the dilatonic scalar
coefficient in the reduced system (\ref{igen}). Given a set of dilaton
vectors for retained field strengths satisfying (\ref{anprod}),
consistency of (\ref{phiperpeq}) with setting $\vec\phi_\perp=0$ requires
\be
\Pi_\perp\cdot\sum_\alpha\vec a_\alpha(F_{\alpha\,[n]})^2 = 0\ .
\ee
This equation requires, for every point $x^{\sst M}$ in spacetime, that
the combination $\sum_\alpha\vec a_\alpha(F_{\alpha\,[n]})^2$ be parallel
to $\vec n$ in the dilaton-vector space. Combining this with the
requirement (\ref{anprod}), one has
\be
\sum_\alpha\vec a_\alpha(F_{\alpha\,[n]})^2 = a\vec
n\sum_\alpha(F_{\alpha\,[n]})^2\ .
\ee
Taking then a dot product of this with $\vec a_\beta$, one has
\be
\sum_\alpha
M_{\beta\alpha}(F_{\alpha\,[n]})^2=a^2\sum_\alpha(F_{\alpha\,[n]})^2\ .
\label{MF2}
\ee
Detailed analysis~~\cite{lp} shows it to be sufficient to consider the
cases where $M_{\alpha\beta}$ is invertible, so by applying
$M^{-1}_{\alpha\beta}$ to (\ref{MF2}), one finds
\be
(F_{\alpha\,[n]})^2 = a^2\sum_\beta
M^{-1}_{\alpha\beta}\sum_\gamma(F_{\gamma\,[n]})^2\ .
\ee
Summing on $\alpha$, one has
\be
a^2=(\sum_{\alpha,\beta}M^{-1}_{\alpha\beta})^{-1}\ ;\label{arel}
\ee
one then defines the retained field-strength combination $F_{[n]}$ so that
\be
(F_{\alpha\,[n]})^2=a^2\sum_\beta M^{-1}_{\alpha\beta}(F_{[n]})^2\ .
\ee

     The only remaining requirement for consistency of the truncation to
the simplified ($g_{\sst MN}$, $\phi$, $F_{[n]}$) system (\ref{igen})
arises from the necessity to ensure that the variation of the ${\cal
L}_{FFA}$ term in (\ref{I_D}) is not inconsistent with setting to zero the
discarded dilatonic scalars and gauge potentials. In general, this imposes
a somewhat complicated requirement. For the purposes of the present review,
however, we shall concentrate on either purely-electric solutions
satisfying the elementary ansatz (\ref{elans}) or purely-magnetic
solutions satisfying the solitonic ansatz (\ref{magans}). As one can see
by inspection, for pure electric or magnetic solutions of these sorts, the
terms that are dangerous for consistency arising from the variation of
${\cal L}_{FFA}$ all vanish. Thus, for such solutions one may safely
ignore the complications of the ${\cal L}_{FFA}$ term. This restriction to
pure electric or magnetic solutions does, however, leave out the very
interesting cases of dyonic solutions that exist in $D=8$ and $D=4$, upon
which we shall comment briefly later on.

     After truncating to the system (\ref{igen}), the analysis proceeds as
in Section~\ref{sec:pbraneans}. It turns out~~\cite{lp} that {\em
supersymmetric} $p$-brane solutions arise when the matrix
$M_{\alpha\beta}$ for the retained $F_{\alpha\,[n]}$ satisfies
\be
M_{\alpha\beta}=4\delta_{\alpha\beta} - {2d\tilde d\over D-2}\ ,
\ee
and the corresponding $\Delta$ value for $F_{[n]}$ is
\be
\Delta={4\over N}\ ,
\ee
where we recall that $N$ is the number of retained field strengths. A
generalization of this analysis leads to a classification of solutions
with more than one independent retained scalar-field combination~~\cite{lp},
but to pursue that would go beyond the focus of the present review, where
we shall limit ourselves to the ``single-scalar'' context.

\section{Multi-center solutions and vertical dimensional
reduction}\label{sec:vertical}

     As we saw above in Section \ref{sec:kkred}, the translation Killing
symmetries of $p$-brane solutions allow a simultaneous interpretation of
such solutions as belonging to several different supergravity theories,
related one to another by Kaluza-Klein dimensional reduction. For the
original single $p$-brane solutions (\ref{pbranesol}), the only available
translational Killing symmetries are those in the worldvolume directions,
which we exploited in describing diagonal dimensional reduction.
One may, however, generalize the basic solutions (\ref{pbranesol}) by
replacing the harmonic function $H(y)$ in (\ref{pbranesol}c) by a
different solution of the Laplace equation (\ref{laplace}). Thus, one can
easily extend the family of $p$-brane solutions to {\em multi-center}
$p$-brane solutions by taking the harmonic function to be
\be
H(y) = 1 + \sum_\alpha {k_\alpha\over|\vec y-\vec y_\alpha|^{\tilde
d}}\hspace{1cm}k_\alpha>0\ .
\label{multiharmonic}
\ee
Once again, the integration constant has been adjusted to make
$H\for_\infty = 1 \leftrightarrow \phi\for_\infty = 0$. The generalized
solution (\ref{multiharmonic}) corresponds to {\em parallel} and {\em
similarly-oriented} $p$-branes, with all charges $\lambda_\alpha = 2\tilde
d k_\alpha/\sqrt\Delta$ required to be positive in order to avoid naked
singularities. The ``centers'' of the individual ``leaves'' of this
solution are at the points $y=y_\alpha$, where $\alpha$ ranges over any
number of centers. The metric and the electric-case antisymmetric tensor
gauge potential corresponding to (\ref{multiharmonic}) are given again in
terms of $H(y)$ by (\ref{pbranesol}a,\ref{csol}). In the solitonic case,
the ansatz  (\ref{magans}) needs to be modified so as to accommodate the
multi-center form of the solution:
\be
F_{m_1\ldots m_n} = -\tilde d^{-1}\epsilon_{m_1\ldots
m_np}\partial_p\sum_\alpha{\lambda_\alpha\over|\vec y-\vec
y_\alpha|^{\tilde d}}\ ,\label{multimagsol}
\ee
which ensures validity of the Bianchi identity just as well
as (\ref{magans}) does. The mass/(unit $p$-volume) density is now
\be
{\cal E} = {1\over2\sqrt\Delta}\sum\alpha\lambda_\alpha\ ,
\ee
while the total electric or magnetic charge is given by
$\ft14\sum\lambda_\alpha$, so the Bogomol'ny bounds (\ref{bogbounds}) are
saturated just as they are for the single-center solutions
(\ref{pbranesol}). Since the multi-center solutions given by
(\ref{multiharmonic}) satisfy the same supersymmetry-preservation
conditions on the metric and antisymmetric tensor as (\ref{pbranesol}), the
multi-center solutions leave the same amount of supersymmetry unbroken as
the single-center solution.

     From a mathematical point of view, the multi-center solutions
(\ref{multiharmonic}) exist owing to the properties of the Laplace
equation (\ref{laplace}). From a physical point of view, however, these
static solutions exist as a result of {\em cancellation} between {\em
attractive} gravitational and scalar-field forces against {\em repulsive}
antisymmetric-tensor forces for the similarly-oriented $p$-brane
``leaves.''

     The multi-center solutions given by (\ref{multiharmonic}) can now be
used to prepare solutions adapted to dimensional reduction in the {\em
transverse} directions. This combination of modifying the solution
followed by dimensional reduction on a transverse coordinate is called
{\em vertical} dimensional reduction~~\cite{vertical} because it relates
solutions vertically on a $D$ versus $d$ plot.\footnote{Similar procedures
have been considered in a number of articles in the literature; see,
{\it e.g.}\ Refs.\cite{kghl}} In order to do this, we need first to
develop a translation invariance in the transverse reduction coordinate.
This can be done by ``stacking'' up identical
$p$ branes using (\ref{multiharmonic}) in a periodic array, {\it i.e.}\ by
letting the integration constants $k_\alpha$ all be equal, and aligning
the ``centers'' $y_\alpha$ along some axis, {\it e.g.}\  the $\hat z$ axis.
Singling out one ``stacking axis'' in this way clearly destroys the overall
isotropic symmetry of the solution, but, provided the centers are all in
a line, the solution will nonetheless remain isotropic in the $D-d-1$
dimensions orthogonal to the stacking axis.  Taking the limit of a
densely-packed infinite stack of this sort, one has
\begin{subeqnarray}
\sum_\alpha{k_\alpha\over|\vec y - \vec y_\alpha|^{\tilde
d}} &\longrightarrow& \int_{-\infty}^{+\infty}{kdz\over(\hat r^2 +
z^2)^{\tilde d/2}} = {\tilde k\over\tilde r^{\tilde d-1}}\\
\hat r^2 &=& \sum_{m=d}^{D-2}y^my^m\\
\hat k &=& {\sqrt\pi k\Gamma(\tilde d - \ft12)\over2\Gamma(\tilde d)}\ ,
\label{stackint}
\end{subeqnarray}
where $\hat r$ in (\ref{stackint}b) is the radial coordinate for the
$D-d-1$ residual isotropic transverse coordinates. After a conformal
rescaling to maintain the Einstein frame for the solution, one can finally
reduce on the coordinate $z$ along the stacking axis.

     After stacking and reduction in this way, one obtains a $p$-brane
solution with the {\em same} worldvolume dimension as the original
higher-dimensional solution that was stacked up. Since
the same antisymmetric tensors are used here to support both the stacked
and the unstacked solutions, and since $\Delta$ is preserved under
dimensional reduction, it follows that vertical dimensional reduction from
$D$ to $D-1$ spacetime dimensions preserves the value of $\Delta$ just
like the diagonal reduction discussed in the previous section. Note
that, under vertical reduction, the worldvolume dimension $d$ is
preserved, but
$\tilde d=D-d-2$ is reduced by one with each reduction step.

\begin{figure}
\epsfbox{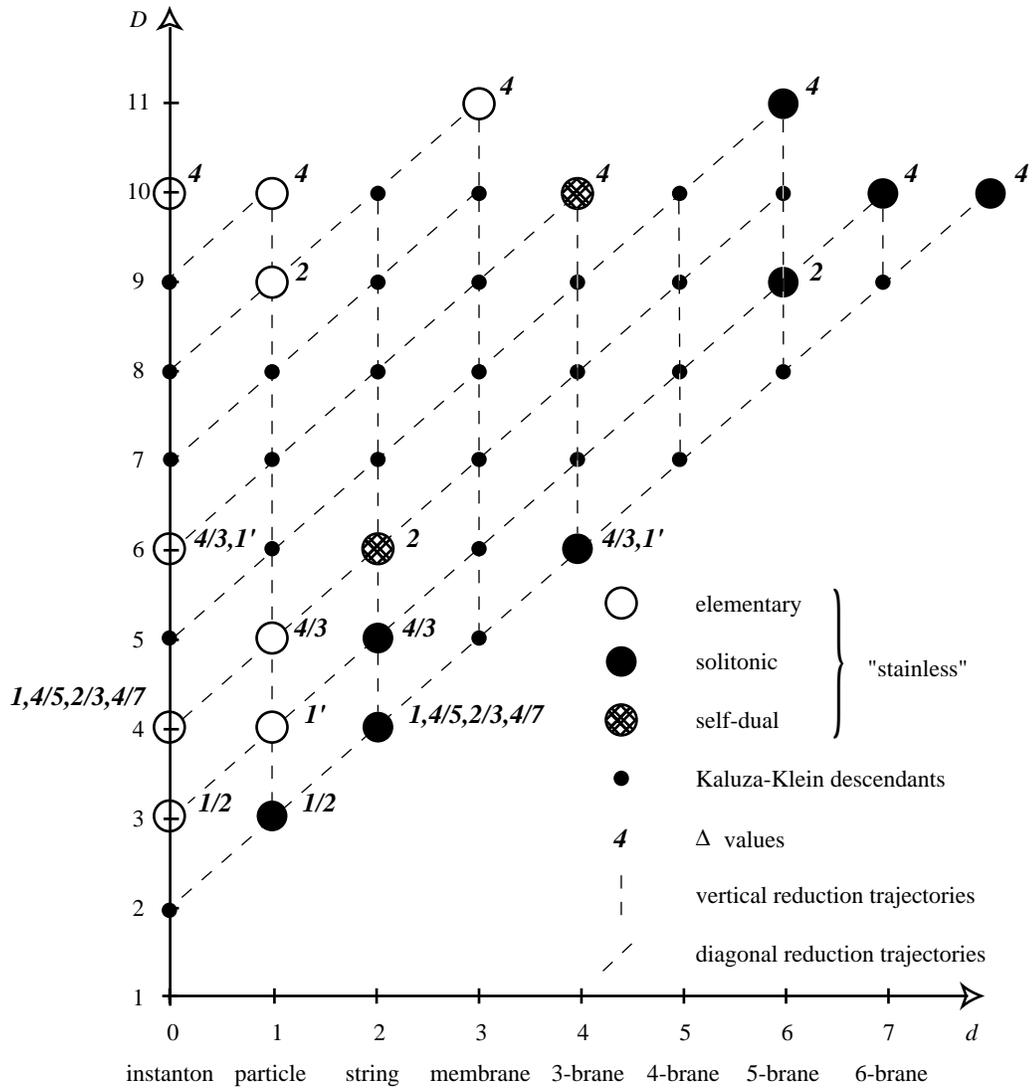}
\caption{Brane-scan of supergravity $p$-brane
solutions ($p\le (D-3)$)\label{fig:nbscan}}
\end{figure}\clearpage

     Combining the diagonal and vertical dimensional reduction
trajectories of ``descendant'' solutions, one finds the general picture
given in the plot of Figure \ref{fig:nbscan}. In this plot of spacetime
dimension $D$ versus worldvolume dimension $d$, reduction families 
emerge from certain basic solutions that cannot be ``oxidized'' to
higher-dimensional solutions, and hence can be called
``stainless'' $p$-branes.\cite{stainless} In Figure \ref{fig:nbscan}, these
solutions are indicated by the large circles, with
the corresponding $\Delta$ values shown
adjacent. The indication of the elementary or solitonic type of solution
relates to solutions of supergravity theories in versions with the lowest
possible choice of rank ($n\le D/2$) for the supporting field strength,
obtainable by appropriate dualization.

\noindent 
\section{Beyond the $(D-3)$-brane
barrier: Scherk-Schwarz reduction and domain walls}\label{sec:domainwalls}

     The process of vertical dimensional reduction described in the
previous section proceeds uneventfully until one makes the reduction from 
a $(D,d=D-3)$ solution to a $(D-1,d=D-3)$ solution.\footnote{Such
solutions, with worldvolume dimension two less than the spacetime
dimension, will be referred to generally as $(D-3)$-branes, irrespective of
whether the spacetime dimension is $D$ or not.} In this step, the integral
(\ref{stackint}) contains an additive divergence and needs to be
renormalized. This is easily handled by putting finite limits $\pm L$ on
the integral, which becomes $\int_{-L}^L d\tilde z(r^2 +\tilde
z^2)^{-1/2}$, and then by subtracting a divergent term $2\ln L$ before
taking the limit $L\rightarrow\infty$. Then the integral gives the expected
$\ln\hat r$ harmonic function appropriate to two transverse dimensions.

     Continuing on down, one may similarly make one more step of 
reduction, from the $(D-1,d=D-3)$ solution obtained above, in an attempt to
create a solution with dimensions $(D-2,d=D-3)$, {\it i.e.} a $(D-2)$-brane,
or {\em domain wall.} In the process of vertical dimensional reduction, this
step again gives rise to an additive divergence: the integral
$\int_{-L}^L dz \ln(y^2 + z^2)$ needs to be renormalized by
subtracting a divergent term $4L(\ln L -1)$. After subsequently performing
the integral, the resulting harmonic $H(y)$ becomes linear in the one
remaining transverse coordinate.

     While the above mathematical procedure of vertical dimensional
reduction to produce a $(D-2)$-brane proceeds apparently without
serious complication, analyzing the physics of the situation needs some
care.\cite{domain} There are three things about the reduction from a
$(D-1,d=D-3)$ solution (a $(D-3)$-brane) to a $(D-2,d=D-3)$ solution (a
$(D-2)$-brane) that require special attention. First, let us note that both
the $(D-3)$ brane and its descendant $(D-2)$-brane have harmonic functions
$H(y)$ that blow up at infinity. For the $(D-3)$-brane, however this is not in 
itself particularly remarkable, because, as one can see by inspection of
(\ref{pbranesol}) for this case, the metric asymptotically tends 
to a locally flat space as $r\rightarrow\infty$, and also in this limit the
dilatonic scalar
$\phi$ and antisymmetric-tensor one-form field strength
\be
F_m = -\epsilon_{mn}\partial_n H\label{oneformfs}
\ee
tend asymptotically to zero. The expression (\ref{oneformfs}) for the field
strength, however, shows that the next reduction step to the $(D-2,d=D-3)$
solution has a significant new feature: in stacking up $(D-3)$-branes
prior to the vertical reduction, producing a linear harmonic function
in the transverse coordinate $y$,
\be
H(y) = \mbox{const.} + my\ ,\label{linharmonic}
\ee
the field strength (\ref{oneformfs}) has a {\em constant} component along
the stacking axis $\leftrightarrow$ reduction direction $z$,
\be
F_z = -\epsilon_{zy}\partial_y H = m\ ,\label{constF}
\ee
that implies an unavoidable dependence\,\footnote{Note that this
vertical reduction from a $(D-3)$-brane to a $(D-2)$-brane is the first
time one is {\em forced} to accept a dependence on the reduction
coordinate $z$; in all higher-dimensional vertical reductions, $z$
dependence can be removed by a gauge transformation. The zero-form gauge
potential in (\ref{zerofmred}) does not have such a gauge symmetry,
however.} of the corresponding zero-form gauge potential on the reduction
coordinate:
\be
A_{[0]}(x,y,z) = mz + \chi(x,y)\ .\label{zerofmred}
\ee

     From a Kaluza-Klein point of view, the unavoidable linear dependence
of a gauge potential on the reduction coordinate given in
(\ref{zerofmred}) appears to be problematic. Throughout this review, we
have dealt only with {\em consistent}) Kaluza-Klein reductions, for which
solutions of the reduced theory are also solutions of the unreduced theory.
Generally, retaining any dependence on a reduction coordinate will lead to
an {\em inconsistent truncation} of the theory: attempting to impose a $z$
dependence of the form given in (\ref{zerofmred}) prior to varying the
Lagrangian will give a result different from that of imposing this
dependence in the field equations after variation.

     Before showing how this consistency problem can sometimes be avoided,
let us consider two other facets of the problem with the vertical
reduction of $(D-3)$-branes. Firstly, the asymptotic metric of a
$(D-3)$-brane is not a globally flat space, but only a locally flat space.
This distinction means that there is in general a deficit solid angle at
transverse infinity, which is related to the total mass density of
the $(D-3)$-brane.\cite{d-3br} This means that any attempt to stack up
$(D-3)$-branes within a standard supergravity theory will soon consume the
entire solid angle at transverse infinity, thus destroying the asymptotic
spacetime by such a construction.

     The second facet of the problem with $(D-2)$-branes in ordinary
supergravity theories is simply stated: starting from the $p$-brane
ansatz (\ref{ansatz},\ref{magans}) and searching for $(D-2)$ branes in
ordinary massless supergravity theories, one simply doesn't find any such
solutions.

     The resolution of all these difficulties happens together, for
in blindly performing a Kaluza-Klein reduction with an ansatz like
(\ref{zerofmred}), one is in fact making a departure from the set of
standard massless supergravity theories. In order to understand this, let
us concentrate on the problem of consistency of the Kaluza-Klein reduction.
As we have seen, consistency of any restriction on the field variables
with respect to the equations of motion means that the restriction may
either be imposed on the field variables in the action prior to variation
to derive the equations of motion, or may be imposed on the field
variables in the equations of motion after variation, with an equal effect.
In that case, solutions obeying the restriction will also be solutions of
the general unrestricted equations of motion.

     The most usual guarantee of consistency in Kaluza-Klein dimensional
reduction is achieved by restricting the field variables to carry zero
charge with respect to some conserved current, {\it e.g.}\ momentum in the
reduction dimension. But this is not the only way in which consistency may
be achieved. In the present case, retaining a linear dependence on the
reduction coordinate as in (\ref{zerofmred}) clearly would produce
an inconsistent truncation if the reduction coordinate were to
appear explicitly in any of the field equations. But this does not imply
that a truncation retaining some dependence on the reduction
coordinate is necessarily inconsistent just because a gauge potential
contains a term linear in that coordinate. Inconsistency of a
Kaluza-Klein truncation occurs when the original unrestricted field
equations imply a condition that is inconsistent with the reduction
ansatz. If a particular gauge potential appears in the action
{\em only through its derivative,} {\it i.e.} through its field strength,
then a consistent truncation may also be achieved provided that the
restriction on the gauge field implies that the field strength is
independent of the reduction coordinate. A zero-form gauge potential on
which such a reduction may be carried out, occurring in the action only
through its derivative, will be referred to as an {\em axion.}

     Requiring axionic field strengths to be independent of the reduction
coordinate amounts to extending the Kaluza-Klein reduction framework to
allow {\em linear} dependence of an axionic zero-form potential on the
reduction coordinate precisely of the form occurring in (\ref{zerofmred}).
So, provided $A_{[0]}$ is an axion, the reduction (\ref{zerofmred}) turns
out to be consistent after all. This extension of the Kaluza-Klein ansatz
is in fact an instance of {\em Scherk-Schwarz} reduction.\cite{ss,bdgpk}
The basic idea of Scherk-Schwarz reduction is to use an Abelian rigid
symmetry of a system of equations, but then to generalize the reduction
ansatz by allowing a linear dependence on the reduction coordinate in the
parameter of this Abelian symmetry. Consistency is guaranteed by
cancellations orchestrated by this Abelian symmetry in field-equation
terms where the parameter does not get differentiated. When it does get
differentiated, it contributes only a term that is itself independent of
the reduction coordinate. In the present case, the Abelian symmetry
guaranteeing consistency of (\ref{zerofmred}) is a simple shift symmetry
$A_{[0]}\rightarrow A_{[0]}+\mbox{const.}$ 

     Unlike the original implementation of the Scherk-Schwarz reduction
idea,\cite{ss} which used an Abelian $U(1)$ phase symmetry acting on
spinors, the Abelian shift symmetry used here commutes with supersymmetry,
and hence the reduction does not spontaneously break supersymmetry.
Instead, gauge symmetries for some of the antisymmetric tensors will be
broken, with a corresponding appearance of mass terms. As with all
examples of vertical dimensional reduction, the $\Delta$ value
corresponding to a given field strength is also preserved. Thus,
$p$-brane solutions related by vertical dimensional reduction, even in the
enlarged Scherk-Schwarz sense, preserve the same amount of unbroken
supersymmetry and have the same value of $\Delta$.    

     It may be necessary to make several redefinitions and integrations by
parts in order to reveal the axionic property of a given zero-form, and
thus to prepare the theory for a reduction of the form (\ref{zerofmred}).
This is most easily explained by an example, so let us consider the first
possible Scherk-Schwarz reduction\,\footnote{A higher-dimensional
Scherk-Schwarz reduction is possible~~\cite{bdgpk} starting from type IIB
supergravity in $D=10$, using the axion appearing in the
\ffrac{$\rm{SL}(2,\R)$}{${\rm SO}(2)$} scalar sector of that theory.} in
the sequence of theories descending from (\ref{D11act}), starting in
$D=9$ where the first axion field appears.\cite{domain} The Lagrangian for
massless $D=9$ maximal supergravity is obtained by specializing the general
dimensionally-reduced action (\ref{I_D}) given in
Section \ref{sec:pbraneans} to this case:
\cramp\bea
{\cal L}_9 &=& \sqrt{-g}\Big[R -\ft12(\partial\phi_1)^2 -\ft12
(\partial\phi_2)^2 -
\ft12 e^{\phi_1+\ft3{\sqrt7}\phi_2} (\partial\chi)^2 -\ft1{48}
e^{\vec a\cdot \vec\phi} (F_{[4]})^2\nonumber\\
&&- \ft12 e^{\vec a_1\cdot \vec\phi} (F_{[3]}^{(1)})^2 -\ft12
e^{\vec a_2\cdot \vec\phi} (F_{[3]}^{(2)})^2 
-\ft14 e^{\vec a_{12}\cdot \vec\phi} (F_{[2]}^{(12)})^2 -\ft14
e^{\vec b_1\cdot\vec\phi} ({\cal F}_{[2]}^{(1)})^2\nonumber\\
&&- \ft14 e^{\vec b_2\cdot \vec\phi} ({\cal F}_{[2]}^{(2)})^2\Big]
-\ft12 \tilde F_{[4]}\wedge \tilde F_{[4]} \wedge A_{[1]}^{(12)} -
\tilde F_{[3]}^{(1)} \wedge \tilde F_{[3]}^{(2)} \wedge A_{[3]}\
,\label{d9lag}
\eea\uncramp
where $\chi={\cal A}_{[0]}^{(12)}$ and $\vec\phi=(\phi_1,\phi_2)$.

     Within the scalar sector $(\vec\phi,\chi)$ of (\ref{d9lag}), the
dilaton coupling has been made explicit; in the rest of the Lagrangian,
the dilaton vectors have the general structure given in
(\ref{dilvecrels},\ref{dilvecprods}). The scalar sector of (\ref{d9lag})
forms a nonlinear $\sigma$-model for the manifold \ffrac{${\rm
GL}(2,\R)$}{${\rm SO}(2)$}. This already makes it appear that one may
identify $\chi$ as an axion available for Scherk-Schwarz reduction.
However, account must still be taken of the Chern-Simons structure lurking
inside the field strengths in (\ref{d9lag}). In detail, the field
strengths are given by
\begin{subeqnarray} 
F_{[4]}&=&\tilde F_{[4]} - \tilde F_{[3]}^{(1)}\wedge {\cal A}_{[1]}^{(1)}
- \tilde F_{[3]}^{(2)}\wedge {\cal A}_{[1]}^{(2)}\nonumber\\
&&\quad\quad +\chi \tilde F_{[3]}^{(1)} \wedge {\cal A}_{[1]}^{(2)} -
\tilde F_{[2]}^{(12)}
\wedge {\cal A}_{[1]}^{(1)}
\wedge {\cal A}_{[1]}^{(2)}\\ 
F^{(1)}_{[3]} &=& \tilde F^{(1)}_{[3]} - \tilde F_{[2]}^{(12)} 
\wedge {\cal A}_{[1]}^{(2)}\\ 
F_{[3]}^{(2)} &=& \tilde F_{[3]}^{(2)} + F_{[2]}^{(12)}\wedge
{\cal A}_{[1]}^{(1)} -
\chi \tilde F^{(1)}_{[3]}\\ 
F_{[2]}^{(12)} &=& \tilde F^{(12)}_{[2]}\quad\quad
{\cal F}_{[2]}^{(1)} = {\cal F}_{[2]}^{(1)} -d\chi\wedge
{\cal A}_{[1]}^{(2)}\\
{\cal F}_{[2]}^{(2)} &=& \tilde {\cal F}_{[2]}^{(2)}
\quad\quad {\cal F}_{[1]}^{(12)} =  d\chi\ ,\label{csterms}
\end{subeqnarray}
where the field strengths carrying tildes are the na\"{\i}ve
expressions without Chern-Simons corrections, {\it i.e.}\ 
$\tilde F_{n]}=dA_{[n-1]}$. Now the appearance of undifferentiated $\chi$
factors in (\ref{csterms}a,c) makes it appear that Scherk-Schwarz
reduction would be inconsistent. However, one may eliminate these
undifferentiated factors by making the field redefinition
\be
A_{[2]}^{(2)}\longrightarrow A_{[2]}^{(2)} +
\chi A_{[2]}^{(1)}\ ,\label{Aredef}
\ee
after which the field strengths (\ref{csterms}a,c) become
\begin{subeqnarray}
F_{[4]}&=&\tilde F_{[4]} - \tilde F_{[3]}^{(1)}\wedge {\cal
A}_{[1]}^{(1)} - \tilde F_{[3]}^{(2)}\wedge {\cal A}_{[1]}^{(2)}\nonumber\\
&&\quad\quad -d\chi \wedge A_{[2]}^{(1)}\wedge {\cal A}_{[1]}^{(2)} -
\tilde F_{[2]}^{(12)}\wedge {\cal A}_{[1]}^{(1)}\wedge {\cal
A}_{[1]}^{(2)}\\
\stepcounter{subequation}
F_{[3]}^{(2)} &=& \tilde F_{[3]}^{(2)} + F_{[2]}^{(12)}\wedge {\cal
A}_{[1]}^{(1)} +
d\chi \wedge A_{[2]}^{(1)}\label{newcsterms}\ ,
\end{subeqnarray}
the rest of (\ref{csterms}) remaining unchanged.

     After making the field redefinitions (\ref{Aredef}), the axion field
$\chi=A_{[0]}^{(12)}$ is ready for application of the Scherk-Schwarz
reduction ansatz (\ref{zerofmred}). The coefficient of the term linear in
the reduction coordinate $z$ has been denoted $m$ because it carries the
dimensions of mass, and correspondingly its effect on the reduced action
is to cause the appearance of mass terms. Applying (\ref{zerofmred}) to
the $D=9$ Lagrangian, one obtains the $D=8$ reduced
Lagrangian~~\cite{domain}
\bea
\rlap{${\cal L}_{8\,{\rm ss}} ~~=~~$}&&\nonumber\\
&&\sqrt{-g}\Big[R -\ft12 
(\partial\phi_1)^2 -\ft12 (\partial\phi_2)^2  -\ft12 
(\partial\phi_3)^2 -
\ft12 e^{\vec b_{12}\cdot \vec \phi}(\partial \chi - m 
{\cal A}_{[1]}^{(3)})^2\nonumber\\
&&-\ft12 e^{\vec b_{13}\cdot \vec \phi}
(\partial {\cal A}_{[0]}^{(13)} + m {\cal A}_{[1]}^{(2)})^2
-\ft12
e^{\vec b_{23}\cdot \vec \phi}(\partial {\cal A}_{[0]}^{(23)})^2 -\ft12
e^{\vec a_{123}\cdot \vec \phi}(\partial A_{[0]}^{(123)})^2
\nonumber\\
&&-\ft1{48} e^{\vec a\cdot\vec \phi} (F_{[4]}- m A_{[2]}^{(1)} \wedge
{\cal A}_{[1]}^{(2)}\wedge {\cal A}_{[1]}^{(3)})^2 
-\ft1{12} e^{\vec a_1\cdot
\vec\phi}(F_{[3]}^{(1)})^2 \nonumber\\
&&-\ft1{12} e^{\vec a_2\cdot
\vec\phi}(F_{[3]}^{(2)}+ m A_{[2]}^{(1)}\wedge {\cal A}_{[1]}^{(3)})^2 
-\ft1{12} e^{\vec a_3\cdot
\vec\phi}(F_{[3]}^{(3)} + m A_{[2]}^{(1)} \wedge {\cal A}_{[1]}^{(2)} 
)^2\nonumber\\
&&-\ft14 e^{\vec a_{12}\cdot \vec \phi}(F_{[2]}^{(12)})^2 
-\ft14 e^{\vec a_{13}\cdot \vec \phi}(F_{[2]}^{(13)})^2
-\ft14 e^{\vec a_{23}\cdot \vec \phi}(F_{[2]}^{(23)}+m A_{[2]}^{(1)})^2
\nonumber\\
&&-\ft14 e^{\vec b_{1}\cdot \vec \phi}({\cal F}_{[2]}^{(1)}
- m {\cal A}_{[1]}^{(2)} \wedge {\cal A}_{[1]}^{(3)})^2
-\ft14 e^{\vec b_{2}\cdot \vec \phi}({\cal F}_{[2]}^{(2)})^2
-\ft14 e^{\vec b_{3}\cdot \vec \phi}({\cal F}_{[2]}^{(3)})^2 \nonumber\\
&& - \ft12 m^2 e^{\vec b_{123}\cdot\vec\phi}\Big] + {\cal L}_{FFA}\ ,
\label{massived8}
\eea
where the dilaton vectors are now those appropriate for $D=8$. It is
apparent from (\ref{massived8}) that the fields ${\cal A}_{[1]}^{(3)}$,
${\cal A}_{[1]}^{(2)}$ and $A_{[2]}^{(1)}$ have now become massive,
absorbing in the process $\chi$, ${\cal A}_{[0]}^{(13)}$ and
$A_{[1]}^{(23)}$. Specifically, these fields are absorbed by making the
following gauge transformations:
\bea
{\cal A}_{[1]}^{(3)}&\longrightarrow & {\cal A}_{[1]}^{(3)} +{1\over m} 
d \chi\nonumber\\
{\cal A}_{[1]}^{(2)}&\longrightarrow &{\cal A}_{[1]}^{(2)} -{1\over m}
d {\cal A}_{[0]}^{(13)}\label{gauge}\\
A_{[2]}^{(1)}&\longrightarrow & A_{[2]}^{(1)} -{1\over m} 
d A_{[1]}^{(23)}\ .\nonumber
\eea
Note that ${\cal A}_{[1]}^{(3)}$ is the Kaluza-Klein vector field
corresponding to this $D=9\rightarrow D=8$ reduction, and that in
becoming massive it consumes the axion $\chi$ on which the Scherk-Schwarz
reduction was performed; this is a general feature of such reductions.
Furthermore, the appearance of the derivative of the reduction axion in
Chern-Simons corrections to other field strengths gives rise to further
spontaneous breakings, in this case giving masses to ${\cal
A}_{[1]}^{(2)}$ and $A_{[2]}^{(1)}$.

     As one descends through the available spacetime dimensions for
supergravity theories, the number of axionic scalars available for a
Scherk-Schwarz reduction step increases. The numbers of axions are given
in the following table:\newpage
\begin{table}[ht]
\caption{Supergravity axions versus spacetime dimension.\label{tab:axions}}
\vspace{0.4cm}
\begin{center}
\begin{tabular}{|l|c|c|c|c|c|c|}
\hline
$D$&9&8&7&6&5&4\\
\hline
$N_{\rm axions}$&1&4&10&20&36&63\\
\hline
\end{tabular}
\end{center}
\end{table}

     Each of these axions gives rise to a {\em distinct} massive
supergravity theory upon Scherk-Schwarz reduction,\cite{domain} and each
of these reduced theories has its own pattern of mass generation. In
addition, once a Scherk-Schwarz reduction step has been performed, the
resulting theory can be further reduced using ordinary Kaluza-Klein
reduction. Moreover, the Scherk-Schwarz and ordinary Kaluza-Klein
processes do not commute, so the number of theories obtained by
the various combinations of Scherk-Schwarz and ordinary dimensional
reduction is cumulative. In addition, there are numerous possibilities of
performing Scherk-Schwarz reduction simultaneously on a number of
axions. As one can see from the $D=9$ example, field redefinitions like
(\ref{Aredef}) need to be made in order to cover axions with derivatives
prior to Scherk-Schwarz reduction, and these redefinitions need to be
consistent if one is to perform such reduction on multiple axions. The
set of axions that can be simultaneously covered by derivatives always
includes the full set of Ramond-Ramond (R-R) sector axions, {\it i.e.}\
those of the forms ${\cal A}_{[0]}^{(1a)}$ or $A_{[0]}^{(abc)}$,
$a,b=2,\ldots,11-D$, which are $2^{9-D}$ in number. There are additional
possibilities involving NS-NS sector axions; for example, in $D=8$, three
out of the four axions shown in Table \ref{tab:axions} may be
simultaneously covered by derivatives: the two R-R axions plus one NS-NS
axion. For further details on this panoply of Scherk-Schwarz reduction
possibilities, we refer the reader to Ref.\cite{domain}

     For our present purposes, the important feature of theories obtained
by Scherk-Schwarz reduction is the appearance of cosmological potential
terms such as the penultimate term in Eq.\ (\ref{massived8}). Such terms
may be considered within the context of our simplified action
(\ref{igen}) by letting the rank $n$ of the field strength take the value
zero. Accordingly, by consistent truncation of (\ref{massived8}) or of one
of the many theories obtained by Scherk-Schwarz reduction in lower
dimensions, one may arrive at the simple Lagrangian
\be
{\cal L}= \sqrt{-g}\Big[R -\ft12\nabla_{\sst M}\phi\nabla^{\sst M}\phi
-\ft12 m^2 e^{a \phi}\Big]\ .\label{coslag}
\ee
Since the rank of the form here is $n=0$, the elementary/electric type
of solution would have worldvolume dimension $d=-1$, which is not
very sensible, but the solitonic/magnetic solution has $\tilde
d=D-1$, corresponding to a $p=D-2$ brane, or {\em domain
wall,} as expected. Relating the parameter $a$ in (\ref{coslag}) to the
reduction-invariant parameter $\Delta$ by the standard formula
(\ref{Delta}) gives $\Delta=a^2-2(D-1)/(D-2)$; taking the corresponding
$p=D-2$ brane solution from (\ref{pbranesol}), one finds\,\footnote{Domain
walls solutions such as (\ref{dwsol}) in supergravity theories were found
for the $D=4$ case in Ref.\cite{c} and a recent review of them has
been given in Ref.\cite{cs}}
\begin{subeqnarray}
ds^2 &=& H^{\ft4{\Delta(D-2)}} \, \eta_{\mu\nu}\, dx^\mu dx^\nu 
+ H^{\ft{4(D-1)}{\Delta(D-2)}} \, dy^2\\
e^\phi &=& H^{2a/\Delta} \ ,\label{dwsol}
\end{subeqnarray}
where the harmonic function $H(y)$ is now a linear function of the single
transverse coordinate, in accordance with (\ref{linharmonic}). The
curvature of the metric (\ref{dwsol}a) tends to zero at large values of
$|y|$, but it diverges if $H$ tends to zero. This latter singularity can be
avoided by taking $H$  to be
\be
H=\mbox{const.} + M|y|\label{dwH}
\ee
where $M=\ft12 m\sqrt\Delta$. With the choice (\ref{dwH}), there is just a
delta-function singularity at the location of the domain wall at $y=0$,
corresponding to the discontinuity in the gradient of $H$.

     The domain-wall solution (\ref{dwsol},\ref{dwH}) has the peculiarity
of tending asymptotically to flat space as $|y|\rightarrow\infty$,
within a theory that does not itself admit flat space as a vacuum solution.
In fact, the theory (\ref{coslag}) does not even admit a
maximally-symmetric solution, owing to the complication of the
cosmological potential. The domain-wall solution (\ref{dwsol},\ref{dwH}),
however, manages to ``cancel'' this potential at transverse infinity,
allowing at least asymptotic flatness for this solution. This brings us
back to the other facets of the consistency problem for vertical
dimensional reduction to produce $(D-2)$-branes as discussed at the
beginning of this section. There is no inconsistency between the existence
of domain-wall solutions like (\ref{dwsol},\ref{dwH}) and the inability to
find such solutions in standard supergravity theories, or with the conical
spacetime character of $(D-3)$-branes because the domain walls exist in a
quite different context of {\em massive} supergravity theories like
(\ref{massived8}) with a vacuum structure different from that of standard
massless supergravities.

\section{Duality symmetries and the classification of
$p$-branes}\label{sec:classification}
\subsection{Supergravity duality symmetries}\label{ssec:dualities}

     As one can see from our discussion of Kaluza-Klein dimensional
reduction in Section \ref{sec:kkred}, progression down to lower dimensions
$D$ causes the number of dilatonic scalars $\vec\phi$ and also the number
of zero-form potentials of 1-form field strengths to proliferate. When one
reaches $D=4$, for example, a total of 70 such spin-zero fields has
accumulated. In $D=4$, the maximal $(N=8)$ supergravity equations of
motion have a linearly-realized $H={\rm SU}(8)$ symmetry; this is also the
automorphism symmetry of the $D=4$, $N=8$ supersymmetry algebra relevant
to the (self-conjugate) supergravity multiplet. In formulating this
symmetry, it is necessary to consider {\em complex} self-dual
and anti-self-dual combinations of the 2-form field strengths, which are
the highest-rank field strengths in $D=4$, higher ranks having been
eliminated by the reduction or by dualization. Using two-component
notation for the $D=4$ spinors, these combinations transform as
$F_{\alpha\beta}^{[ij]}$ and $\bar F_{\dot\alpha\dot\beta\,[ij]}$,
$i,j=1,\ldots,8$, {\it i.e.}\ as a complex 28-dimensional dimensional
representation of ${\rm SU}(8)$. Since this complex representation can be
carried only by the complex field-strength combinations and not by the
1-form gauge potentials, it cannot be locally formulated at the level of
the gauge potentials or of the action, where only an ${\rm SO}(8)$
symmetry is apparent.

     Taking all the spin-zero fields together, one finds that they form a
rather impressive nonlinear $\sigma$-model on a 70-dimensional manifold.
Anticipating that this manifold must be a coset space with $H={\rm SU}(8)$
as the linearly-realized denominator group, Cremmer and Julia~~\cite{cj}
deduced that it had to be the manifold \ffrac{${\rm E}_{7(+7)}$}{${\rm
SU}(8)$}; since the dimension of ${\rm E}_7$ is 133 and that of ${\rm
SU}(8)$ is 63, this gives a 70-dimensional manifold. Correspondingly, a
nonlinearly-realized ${\rm E}_{7(+7)}$ symmetry also appears as an
invariance of the $D=4$, $N=8$ maximal supergravity equations of motion.
Such nonlinearly-realized symmetries of supergravity theories have always
had a somewhat mysterious character. They arise in part out of general
covariance in the higher dimensions, from which supergravities arise by
dimensional reduction, but this is not enough: such symmetries act
transitively on the $\sigma$-model manifolds, mixing fields arising 
both from the metric and from the reduction of the $D=11$ 3-form potential
$A_{[3]}$ in (\ref{D11act}).

     In dimensions $4\le D\le9$, maximal supergravity has the
sets of $\sigma$-model nonlinear $G$ and linear $H$ symmetries shown in
Table \ref{tab:sugrasyms}. In all cases, the spin-zero fields take their
values in ``target'' manifolds \ffrac{$G$}{$H$}. Just as the asymptotic
value at infinity of the metric defines the reference, or ``vacuum''
spacetime with respect to which integrated charges and energy/momentum are
defined, so do the asymptotic values of the spin-zero fields define the
``scalar vacuum.'' These asymptotic values are referred to as the {\em
moduli} of the solution. In string theory, these moduli acquire
interpretations as the {\em coupling constants} and vacuum {\em
$\theta$-angles} of the theory. Once these are determined for a given
``vacuum,'' the {\em classification symmetry} that organizes the distinct
solutions of the theory into families with the same energy must be a
subgroup of the {\em little group}, or {\em isotropy group,} of the
vacuum. In ordinary General Relativity with asymptotically flat
spacetimes, the analogous group is the spacetime Poincar\'e group times
the appropriate classifying symmetry for internal symmetries, such as the
group of rigid ({\it i.e.}\ constant-parameter) Yang-Mills gauge
transformations.

     The isotropy group of any point on a coset manifold \ffrac{$G$}{$H$}
is just $H$, so this is the classical ``internal'' classifying symmetry for
supergravity.

\begin{table}[h]
\caption{Supergravity $\sigma$-model symmetries.\label{tab:sugrasyms}}
\vspace{0.4cm}
\begin{center}
\begin{tabular}{|c|c|c|}
\hline
$D$&$G$&$H$\\
\hline
9&${\rm GL}(2,\R)$&${\rm SO}(2)$\\
\hline
8&${\rm SL}(3,\R)\times{\rm SL}(2,\R)$&${\rm SO}(3)\times{\rm
SO}(2)$\\
\hline
7&${\rm SL}(5,\R)$&${\rm SO}(5)$\\
\hline
6&${\rm SO}(5,5)$&${\rm SO}(5)\times{\rm SO}(5)$\\
\hline
5&${\rm E}_{6(+6)}$&${\rm USP}(8)$\\
\hline
4&${\rm E}_{7(+7)}$&${\rm SU}(8)$\\
\hline
\end{tabular}
\end{center}
\end{table}

\subsection{An example of duality symmetry: $D=8$ supergravity}
\label{ssec:d8sugra}

     In maximal $D=8$ supergravity, one sees from Table
\ref{tab:sugrasyms} that $G={\rm SL}(3,\R)\times{\rm SL}(2,\R)$ and the
isotropy group is $H={\rm SO}(3)\times{\rm SO}(2)$. We have an $(11-3=8)$
vector of dilatonic scalars as well as a singlet $F_{[1]}^{ijk}$ and a
triplet ${\cal F}_{[1]}^{ij}$ $(i,j,k = 1,2,3)$ of 1-form field strengths
for zero-form potentials. Taken all together, we have a manifold of
dimension 7, which fits in precisely with the dimension of the
\ffrac{$({\rm SL}(3,\R)\times{\rm SL}(2,\R))$}{$({\rm SO}(3)\times{\rm
SO}(2))$} coset-space manifold: $8+3-(3+1)=7$.

     Owing to the direct-product structure, we may for the time being
eliminate the 5-dimensional \ffrac{${\rm SL}(3,\R)$}{${\rm SO}(3)$}
sector and consider for simplicity just the 2-dimensional \ffrac{${\rm
SL}(2,\R)$}{${\rm SO}(2)$} sector. Here is the relevant part of the
action:\,\cite{ilpt}
\bea
I_8^{{\rm SL}(2)} &=& \int d^8x\sqrt{-g}\Big[R-\ft12\nabla_{\sst
M}\sigma\nabla^{\sst M}\sigma - \ft12e^{-2\sigma}\nabla_{\sst
M}\chi\nabla^{\sst M}\chi\nonumber\\
&&\hspace{3cm} -
{1\over2\cdot4!}e^\sigma(F_{[4]})^2 - {1\over2\cdot4!}\chi
F_{[4]}{^\ast}F_{[4]}\Big]\label{d8sl2}
\eea
where ${^\ast}F^{\sst MNPQ}=1/(4!\sqrt{-g})\epsilon^{{\sst
MNPQ}x_1x_2x_3x_4}F_{x_1x_2x_3x_4}$ (the $\epsilon^{[8]}$ is a density, so
purely numerical).

     On the scalar fields $(\sigma,\chi)$, the ${\rm SL}(2,\R)$ symmetry
acts as follows: let $\lambda=\chi+\im e^\sigma$; then 
\be
\Lambda=\pmatrix{a&b\cr c&d}
\ee
with $ab-cd=1$ is an
element of ${\rm SL}(2,\R)$ and acts on $\lambda$ by the
fractional-linear transformation
\be
\lambda\longrightarrow {a\lambda + b\over c\lambda + d}\ .\label{fractlin}
\ee

     The action of the ${\rm SL}(2,\R)$ symmetry on the 4-form field
strength gives us an example of a symmetry of the equations of motion that
is not a symmetry of the action. The field strength $F_{[4]}$ forms an
${\rm SL}(2,\R)$ {\em doublet} together with
\be
G_{[4]}=e^\sigma{^\ast}F_{[4]} - \chi F_{[4]}\ ,
\ee
{\it i.e.,}
\be
\pmatrix{F_{[4]}\cr G_{[4]}} \longrightarrow (\Lambda^{\rm
T})^{-1}\pmatrix{F_{[4]}\cr G_{[4]}}\ .\label{F4transf}
\ee
One may check that these transform the $F_{[4]}$ field equation
\be
\nabla_{\sst M}(e^\sigma F^{\sst MNPQ} + \chi {^\ast}F^{\sst MNPQ}) = 0
\label{F4eq}
\ee
into the corresponding Bianchi identity,
\be
\nabla_{\sst M}{^\ast}F^{\sst MNPQ} = 0\ .\label{F4bianchi}
\ee
Since the field equations may be expressed purely in terms of $F_{[4]}$,
we have a genuine symmetry of the field equations in the transformation
(\ref{F4transf}), but since this transformation cannot be expressed
locally in terms of the gauge potential $A_{[3]}$, this is not a
local symmetry of the action. The transformation
(\ref{fractlin},\ref{F4transf}) is a
$D=8$ analogue of an ordinary Maxwell duality transformation in the
presence of scalar fields. Accordingly, we shall refer to the supergravity
$\sigma$-model symmetries generally as duality symmetries.

     The $F_{[4]}$ field strength of the $D=8$ theory supports
elementary/electric $p$-brane solutions with $p=4-2=2$, {\it i.e.}\
membranes, which have a $d=3$ dimensional worldvolume. The corresponding
solitonic/magnetic solutions in $D=8$ have worldvolume dimension $\tilde
d=8-3-2=3$ also. So in this case, $F_{[4]}$ supports {\em both} electric
and magnetic membranes. It is also possible in this case to have solutions
generalizing the purely electric or magnetic solutions that we have
considered to solutions that carry both types of charge, {\it i.e.}\  {\em
dyons}.\cite{ilpt} This possibility is also reflected in the
combined Bogomol'ny bound for this situation, which generalizes the
single-charge bounds (\ref{bogbounds}):
\be
{\cal E}^2 \ge e^{-\sigma_\infty}(U + \chi_\infty V)^2 +
e^{\sigma_\infty} V^2\ ,\label{d8bogbound}
\ee
where $U$ and $V$ are the electric and magnetic charges and
$\sigma_\infty$ and $\chi_\infty$ are the moduli, {\it
i.e.}\ the constant asymptotic values of the scalar fields $\sigma(x)$ and
$\chi(x)$.\footnote{In comparing (\ref{d8bogbound}) to the single-charge
bounds (\ref{bogbounds}), one should take note that for $F_{[4]}$ in
(\ref{d8sl2}) we have $\Delta=4$, so $2/\sqrt\Delta = 1$.} The bound
(\ref{d8bogbound}) is itself ${\rm SL}(2,\R)$ invariant, provided that one
transforms in general the moduli $(\sigma_\infty,\chi_\infty)$ (according
to (\ref{fractlin})) as well as the charges $(U,V)$. For the simple case
with
$\sigma_\infty=\chi_\infty=0$ that we have mainly considered, the bound
(\ref{d8bogbound}) reduces to ${\cal E}^2\ge U^2 + V^2$, which is
invariant under an obvious isotropy group $H={\rm SO}(2)$.

\subsection{Charge quantization}\label{ssec:chargequantization}

     So far, we have discussed the structure of $p$-brane solutions at a
purely classical level. At this level, a given supergravity theory can
have a continuous spectrum of electrically and magnetically charged
solutions with respect to any one of the $n$-form field strengths that can
support such solutions. At the quantum level, however, an important
restriction on this spectrum of solutions enters into force. The
Dirac-Schwinger-Zwanziger (DSZ) quantization conditions for particles with
electric or magnetic charges or for the charges of dyonic particles
generalize to $p$-branes as well.\cite{nt} For the simplest case of
vanishing moduli, {\it e.g.}\ $\sigma_\infty=\chi_\infty=0$ for our $D=8$
example, and after suitable normalization,\footnote{For the case at
hand,\cite{ilpt} one has
$q={1\over\Omega_4}\int_{\partial M_{5{\sst T}}}G$,
$p={1\over\Omega_4}\int_{\partial M_{5{\sst T}}}F$.} the electric and
magnetic charge density numbers $(q,p)$ with respect to a given field
strength $F_{[n]}$ are required to satisfy the relation
\be
(qp'-q'p) \in \Z\ ,\label{chgquant}
\ee
where $(q,p)$ and $(q',p')$ are the charge density numbers of any two
solutions in the spectrum. If, in addition, one assumes the existence of a
singly-charged {\em purely electric} solution with charge density numbers
$(1,0)$, then the allowed charge density numbers are constrained to lie on
an integer {\em charge lattice}: $q,p \in \Z$.

     The quantum-level restriction of allowed charges to a charge lattice
has an impact on the allowed symmetry transformations, since electric and
magnetic charges are acted upon by supergravity duality symmetries; {\it
c.f.}\ (\ref{F4transf}). For the simple case of vanishing scalar moduli,
this restricts the transformations to those respecting an integer charge
lattice. In the $D=8$ example, this restricts the allowed ${\rm
SL}(2,\R)$ matrices to be {\em integer-valued}, thus restricting ${\rm
SL}(2,\R)$ to ${\rm SL}(2,\Z)$. In the general case, the supergravity
duality group ($\sigma$-model symmetry group) $G$ given in Table
\ref{tab:sugrasyms} is restricted to $G(\Z)$ in an analogous fashion. In
the case of the Cremmer-Julia duality groups in lower dimensions, there is
an appropriate definition~~\cite{ht} of discretized duality groups like
$E_{7(+7)}(\Z)$ as the set of ${\rm SP}(56,\Z)$ matrices that preserve the
$E_7$ quadratic invariant.

\subsection{Counting $p$-branes}

     As we have seen at the classical level, the classifying symmetry for
solutions in a given scalar vacuum, specified by the values of the scalar
moduli, is the linearly-realized isotropy symmetry $H$ given in Table
\ref{tab:sugrasyms}. When one takes into account the DSZ quantization
condition, this classifying symmetry also becomes restricted to a discrete
group, which clearly must be a subgroup of the corresponding $G(\Z)$, so
in general one seeks to identify the group $G(\Z)\cap H$. The value of
this intersection is modulus-dependent, showing that the homogeneity of
the \ffrac{$G$}{$H$} coset space is broken at the quantum level by the
quantization condition. Classically, of course, the particular point on
the vacuum manifold \ffrac{$G$}{$H$} corresponding to the scalar moduli can
be changed by application of a transitively-acting $G$ transformation, for
example with group element $g$. Correspondingly, the isotropy subgroup $H$
moves by conjugation with $g$,
\be
H\longrightarrow gHg^{-1}\ .\label{Hconj}
\ee
The discretized duality group $G(\Z)$ also moves by conjugation, but in
the opposite way,
\be
G(\Z)\longrightarrow g^{-1}G(\Z)g\ ,\label{GZconj}
\ee
so the intersection $G(\Z)\cap H$ takes different values depending on the
moduli. For comparison, in ordinary Maxwell theory, one only has a true
duality symmetry when the electric charge takes the value unity (in
appropriate units), since the duality transformation maps $e\rightarrow
e^{-1}$. Thus, the value $e=1$ is a distinguished value.

     The distinguished point on the scalar vacuum manifold for general
supergravity theories is the one where all scalar moduli vanish. This is
the point where $G(\Z)\cap H$ takes its maximal value. Let us return to
our $D=8$ example to identify what this group is. In that case, for the
scalars $(\sigma,\chi)$, we may write out the transformation in detail
using (\ref{fractlin}):
\bea
e^{-\sigma}&\longrightarrow&(d+c\chi)^2e^{-\sigma} + c^2e^\sigma\nonumber\\
\chi e^{-\sigma}&\longrightarrow& (d + c\chi)(b + a\chi)e^{-\sigma} +
ace^\sigma\ .\label{sigmachitransfs}
\eea
Requiring $a,b,c,d \in \Z$ and also that the point
$\sigma_\infty=\chi_\infty=0$ be left invariant, we find only two
transformations: the identity and $a=d=0, b=-1, c=1$, which maps
\bea
e^{-\sigma}&\longrightarrow&e^\sigma+\chi^2e^{-\sigma}\nonumber\\
\chi e^{-\sigma}&\longrightarrow&-\chi e^{-\sigma}\
.\label{discretesigmachitransfs}
\eea
Thus, for our truncated system, we find just an $S_2$ discrete symmetry as
the quantum isotropy subgroup of ${\rm SL}(2,\Z)$ at the distinguished
point on the vacuum manifold. This $S_2$ is the natural analogue of the
$S_2$ symmetry that appears in Maxwell theory when $e=1$.

     In order to aid in identifying the pattern behind the above $D=8$
example, suppose that the zero-form gauge potential $\chi$ is small, and
consider the $S_2$ transformation to lowest order in $\chi$. To this
order, the transformation just flips the signs of $\sigma$ and $\chi$.
Acting on the field strengths $(F_{[4]},G_{[4]})$, one finds
\be
(F_{[4]},G_{[4]})\longrightarrow (-G_{[4]},F_{[4]})\ .\label{infchitransf}
\ee
One may again check (in fact to all orders, not just to lowest order in
$\chi$) that (\ref{infchitransf}) maps the field equation for $F_{[4]}$
into the corresponding Bianchi identity:
\be
\nabla_{\sst M}(e^\sigma F^{\sst MNPQ} + \chi{^\ast}F^{\sst
MNPQ})\longrightarrow - \nabla_{\sst M}{^\ast}F^{\sst MNPQ}\ .
\ee
Considering the $S_2$ transformation to lowest order in the zero-form
$\chi$ has the advantage that the sign-flip of $\phi$ may be ``impressed''
upon the $\vec a$ dilaton vector for $F_{[4]}$: $\vec a\rightarrow -\vec
a$. The general structure of such $G(\Z)\cap H$ transformations will be
found by considering the impressed action of this group on the dilaton
vectors.

     Now consider the \ffrac{${\rm SL}(3,\R)$}{${\rm SO}(3)$} sector of the
$D=8$ scalar manifold, again with the moduli set to the distinguished point
on the $\sigma$-model manifold. To lowest order in zero-form gauge
potentials, the action of ${\rm SL}(3,\Z)\cap H$ may similarly by impressed
upon the 3-form dilaton vectors, causing in this case a {\em permutation}
of the $\vec a_i$, generating for the $D=8$ case overall the discrete group
$S_3\times S_2$. Now that we have a bit more structure to contemplate, we
can notice that the $G(\Z)\cap H$ transformations {\em leave the $(\vec
a,\vec a_i)$ dot products invariant.}\cite{weyl}

     The invariance of the dilaton vectors' dot products prompts one to
return to the algebra (\ref{dilvecprods}) of these dot products and see
what else we may recognize in it. Noting that the duality groups given in
Table \ref{tab:sugrasyms} for the higher dimensions $D$ involve
${\rm SL}(N,\R)$ groups, we recall that the {\em weight vectors} $\vec
h_i$ of the fundamental representation of ${\rm SL}(N,\R)$ satisfy
\be
\vec h_i\cdot\vec h_j = \delta_{ij}-{1\over N}\ ,\qquad\qquad \sum_{i=1}^N
\vec h_i = 0\ .\label{weightprods}
\ee
These relations are precisely those satisfied by ${\pm 1\over\sqrt2}\vec
a$ and ${1\over\sqrt2}\vec a_i$, corresponding to the cases $N=2$ and
$N=3$. This suggests that the action of the maximal $G(\Z)\cap H$
group ({\it i.e.}\ that for scalar moduli set to the distinguished point on
the $\sigma$-model manifold) may be identified in general with the symmetry
group of the set of fundamental weights for the corresponding supergravity
duality group $G$ as given in Table \ref{tab:sugrasyms}. The symmetry
group of the fundamental weights is the {\em Weyl group}~~\cite{weyl} of
$G$, so the action of the maximal $G(\Z)\cap H$ $p$-brane classifying
symmetry becomes identified with that of the Weyl group of $G$.

     As one proceeds through the lower-dimensional cases, where the
supergravity symmetry groups shown in Table \ref{tab:sugrasyms} grow in
complexity, the above pattern persists:\,\cite{weyl} in all cases, the
action of the maximal classifying symmetry $G(\Z)\cap H$ may be identified
with the Weyl group of $G$. This is then the group that {\em counts} the
distinct p-brane solutions\,\footnote{Of course, these solutions must also
fall into supermultiplets with respect to the unbroken supersymmetry; the
corresponding supermultiplet structures have been discussed in
Ref.\cite{dr}} of a given mass density (\ref{admenerg}), subject to the
DSZ quantization condition and referred to the distinguished point on the
manifold of scalar moduli. For example, in $D=7$, where from Table
\ref{tab:sugrasyms} one sees that $G={\rm SL}(5,\R)$ and $H={\rm SO}(5)$,
one finds that the action of $G(\Z)\cap H$ is equivalent to that of the
discrete group $S_5$, which is the Weyl group of ${\rm SL}(5,\R)$. In the
lower-dimensional cases shown in Table \ref{tab:sugrasyms}, the  discrete
group $G(\Z)\cap H$ becomes less familiar, and is most simply described
as the Weyl group of $G$.

\begin{table}[h]
\caption{Examples of $p$-brane Weyl-group
multiplicities\label{tab:weylmults}}
\vspace{0.4cm}
\begin{center}
\begin{tabular}{|c|c||c|c|c|c|c|c|c|}
\cline{3-9}
\multicolumn{2}{c|}{}&\multicolumn{7}{|c|}{$D$}\\
\cline{3-9}\hline
$F_{[n]}$&$\Delta$&10&9&8&7&6&5&4\\
\hline\hline
$F_{[4]}$&4&1&1&2&\multicolumn{4}{c|}{}\\
\hline
$F_{[3]}$&4&1&2&3&5&10&\multicolumn{2}{c|}{}\\
\hline
&4&1&1+2&6&10&16&27&56\\
\cline{2-9}
$F_{[2]}$&2&&2&6&15&40&135&756\\
\cline{2-9}
&\ffrac43&\multicolumn{5}{c|}{}&45&2520\\
\hline
&4&&2&8&20&40&72&126\\
\cline{2-9}
$F_{[1]}$&2&\multicolumn{2}{c|}{}&12&60&280&1080&3780\\
\cline{2-9}
&\ffrac43&\multicolumn{4}{c|}{}&480&4320&30240+2520\\
\hline
\end{tabular}
\end{center}
\end{table}

     From the above analysis of the Weyl-group duality multiplets, one may
tabulate~~\cite{weyl} the multiplicities of $p$-branes residing at each
point of the plot given in Figure \ref{fig:nbscan}. For supersymmetric
$p$-branes arising from a set of $N$ participating field strengths $F_{[n]}$,
corresponding to $\Delta=4/N$ for the dilatonic scalar coupling, one finds the
multiplicities shown in Table \ref{tab:weylmults}. By combining these
duality multiplets together with the diagonal and vertical dimensional
reduction families discussed in Sections \ref{sec:kkred} and
\ref{sec:vertical}, the full set of $p\le(D-3)$ branes shown in
Figure \ref{fig:nbscan} becomes ``welded'' together into one overall
symmetrical structure.

\section{Concluding remarks}\label{sec:concl}

     In this review, we have focused on the basic structure of $p$-brane
solutions to supergravity theories. Many aspects of this story invite
further elaboration. For simplicity, we have concentrated on solutions
involving just one independent dilatonic-scalar combination corresponding
to the decomposition (\ref{phidecomp}); this can, however, be generalized
to multi-scalar solutions as shown in Ref.\cite{lp2} We have also
concentrated on fully-isotropic solutions to the transverse-space Laplace
equation (\ref{laplace}); these may be interpreted as single $p$-brane
hyperplanes embedded into the ambient $D$-dimensional spacetime. This
construction also can be generalized, by allowing the harmonic function
$H(y)$ to have less than full isotropicity, giving ``intersecting
$p$-brane'' solutions~~\cite{intersecting} for which the separation into
``worldvolume'' and ``transverse'' directions varies as one moves about at
infinity, with only a subset of these directions being ``overall
worldvolume'' or ``overall transverse,'' the rest having a ``relative''
worldvolume or transverse character. Another aspect of the $p$-brane story
that we have not covered here is the sense in which $\Delta=4/N$ solutions
for $N\ge2$ may be considered to be ``bound states'' of $\Delta=4$
solutions at threshold, {\it i.e.}\ with zero binding
energy.\cite{kklp,dr} A further generalization of this story is to
multiple $p$-brane solutions linked by branes of lesser
dimensionality,\cite{surgery} a construction which has been termed ``brane
surgery.''

     At the classical level that we have confined ourselves to in this
review, the singularity structures of $p$-brane supergravity solutions
vary significantly as one moves around the brane-scan of Figure
\ref{fig:nbscan}. Solutions involving scalar fields are singular at the
horizon; extremal solutions without scalars can be continued inside the
horizon, either yielding an overall non-singular spacetime such as the
$D=11$ five-brane shown in Figure \ref{fig:D11cpmag}, or leading to
timelike interior singularities such as that for the $D=11$ membrane shown
in Figure \ref{fig:D11elcp}. The fact that all of the non-extremal
``black-brane'' solutions for $p\ge1$ are singular at their inner horizons
makes a curious contrast with the non-singular horizons of the extremal
cases. Clearly, much remains to be understood about this subject, which
may shed further light on the sense in which the extremal BPS $p$-branes
may be considered to be fundamental excitations of the underlying quantum
theory, an interpretation that would not seem appropriate for the
non-extremal cases. If the analogy to the standard Reissner-Nordstrom
solution is instructive, it would appear that most of the details
of a solution inside the horizon would not have an effect visible in the
exterior solution; this may be expected to be encoded in generalizations of the
standard ``no-hair'' theorems for black holes.

     When there are timelike singularities in $p$-brane solutions, one is
clearly invited to try to couple in a source. In the present review, we
have not engaged in this discussion for reasons of simplicity, but, of
course, a considerable amount is known about the structure of
such $p$-brane worldvolume actions.\cite{dkl,mjd} The worldvolume actions
have been known for some time for those $p$-branes supported by NS-NS
sector field strengths with $\Delta=4$; these constructions follow the
pattern of the original $D=11$ supermembrane action.\cite{bst} The main
difficulty is to square the $p$-brane's partial supersymmetry breaking with
the original full supersymmetry and Lorentz symmetry of its parent supergravity
theory, by the construction of a ``$\kappa$-symmetric'' worldvolume action. The
$\kappa$ symmetry achieves an embedding of a partially-nonlinear realisation of
supersymmetry into a fully-linear realisation, by the introduction of redundant
fermionic gauge degrees of freedom. There is currently much effort being
devoted to finding $\kappa$-symmetric worldvolume actions for the remaining
unsolved cases involving R-R sector antisymmetric-tensor fields.\cite{kappa}

     Of course, the real fascination of this whole subject lies in its
connection to emerging understandings in string theory/quantum gravity,
and in the possibility of determining some of the structure of that theory
by knowledge of its fundamental/solitonic state spectrum, perhaps {\it via}
an ``inverse scattering'' analogy to methods that have been very powerful
in the study of integrable models.

\section*{Acknowledgments}
The author would like to acknowledge helpful conversations with Marcus
Bremer, Fran\c{c}ois Englert, Hong L\"u, George Papadopoulos, Chris Pope,
Paul Townsend, Walter Troost and Antoine Van Proeyen. The author would like
to thank the Institute for Theoretical Physics at K.U. Leuven, the I.C.T.P.
and S.I.S.S.A. in Trieste, and the Yukawa Institute of the University of
Kyoto for hospitality at various times during the preparation of this
review. This work was supported in part by the Commission of the European
Communities under contracts SCI*-CT92-0789 and ERBFMRX-CT96-0045.

\end{document}